\documentclass[%
reprint,
superscriptaddress,
nofootinbib,
 amsmath,amssymb,
nolongbibliography
]{revtex4-2}
\usepackage{bbold}
\usepackage{graphicx}
\usepackage{dcolumn}
\usepackage{bm}
\usepackage[colorlinks=true, allcolors=blue]{hyperref}
\usepackage{xcolor}

\allowdisplaybreaks

\usepackage{enumerate}
\usepackage{physics}
\usepackage{amsmath}
\usepackage{amssymb}
\usepackage{mathrsfs}
\usepackage{subfigure}
\usepackage{soul}
\allowdisplaybreaks

\begin{document}


\title{Instantaneous tunneling time within the theory of time-of-arrival operators}

\author{Philip Caesar Flores}
\email{flores@mbi-berlin.de}
\affiliation{Max-Born-Institute, Max-Born Straße 2A, 12489 Berlin, Germany}
\author{Dean Alvin Pablico}
\email{dlpablico@up.edu.ph}
\affiliation{%
	National Institute of Physics, University of the Philippines Diliman, 1101 Quezon City, Philippines
}
\affiliation{%
University of Northern Philippines, 2700 Vigan City, Ilocos Sur, Philippines
}%
\author{Eric Galapon}%
\email{eagalapon@up.edu.ph}
\affiliation{%
	National Institute of Physics, University of the Philippines Diliman, 1101 Quezon City, Philippines
}

\date{\today}

\begin{abstract}
It was shown in \href{https://journals.aps.org/prl/abstract/10.1103/PhysRevLett.108.170402}{\textit{Phys. Rev. Lett.}, \textbf{108} 170402 (2012)}, that quantum tunneling is instantaneous using a time-of-arrival (TOA) operator constructed by Weyl quantization of the classical TOA. However, there are infinitely many possible quantum images of the classical TOA, leaving it unclear if one is uniquely preferred over the others. This raises the question on whether instantaneous tunneling time is simply an artifact of the chosen ordering rule. Here, we demonstrate that tunneling time vanishes for all possible quantum images of the classical arrival time, irrespective of the ordering rule between the position and momentum observables. The result still holds for TOA-operators that are constructed independent of canonical quantization, while still imposing the correct algebra defined by the time-energy canonical commutation relation. 
\end{abstract}. 

\maketitle


\section{Introduction}

\begin{figure}[t!]
\centering
\begin{subfigure}
    \centering
    \includegraphics[width=0.22\textwidth]{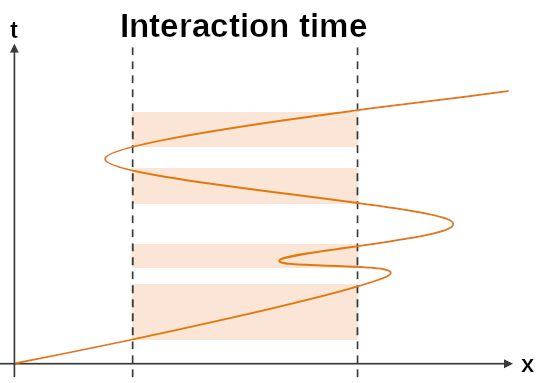}
\end{subfigure} \hfil
\begin{subfigure}
    \centering
    \includegraphics[width=0.22\textwidth]{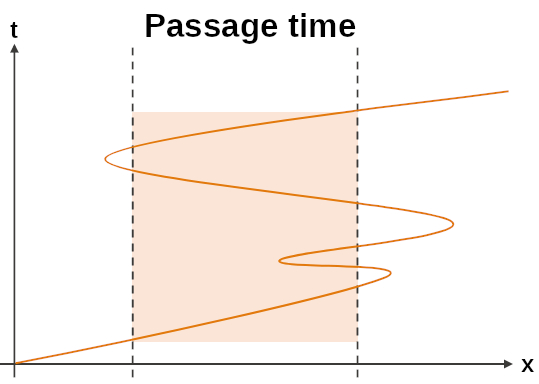}
\end{subfigure} \hfil
\caption{Operational definition of interaction and passage time in terms of Feynman paths. The interaction time is the total time the particle spends inside the barrier while passage time is the time difference between the first entry and last exit (adopted from Ref. \cite{yamada2004unified}).}
\label{fig:feynmanpath}
\end{figure}

The quantum tunneling time problem has been a long standing problem in quantum mechanics \cite{maccoll1932note,hartman1962tunneling} because time is not an observable in standard quantum mechanics due to Pauli's no-go theorem, which asserts the non-existence of a self-adjoint time operator canonically conjugate to a semi-bounded Hamiltonian \cite{pauli1933handbuch}. As such, quantum mechanics offers no canonical formalism on how to answer questions involving time durations. Nevertheless, the prevalent parametric treatment of time has lead to several definitions of the tunneling time, such as Wigner phase time \cite{wigner1955lower}, B\"{u}ttiker-Landauer time \cite{buttiker1982traversal}, Larmor time \cite{baz1966lifetime,rybachenko1967time,buttiker1983larmor}, Pollak-Miller time \cite{pollak1984new}, and dwell time \cite{smith1960lifetime}, among many others  \cite{sokolovski1987traversal,de2002time,winful2006tunneling,imafuku1997effects,brouard1994systematic,jaworski1988time,leavens1989dwell,hauge1987transmission,yamada2004unified,PhysRevLett.127.133001,araujo2024space,de2024traveling}. The validity of these various proposals remains debated, and it is still unclear how they all relate to one another \cite{winful2003nature,PhysRevLett.127.133001}. However, it was shown in Ref. \cite{yamada2004unified} that all these `times' can be categorized as either an \textit{interaction time} or \textit{passage time}, with the distinction between the two depending on the time taken by a Feynman path to traverse the barrier region, as illustrated in Fig. \ref{fig:feynmanpath}. 

A seminal experiment done by Steinberg \textit{et al.} in 1993 which compared the time-of-arrival (TOA) of two entangled photons in the presence and absence of a potential barrier \cite{steinberg1993measurement} showed that the presence of a barrier leads to earlier arrival times that are consistent with the Wigner phase time. However, it did not settle the tunneling time problem because as it turns out, different experimental setups may measure different tunneling times \cite{chiao1997vi}. Moreover, recent attoclock experiments have reported instantaneous tunneling time \cite{eckle2008attosecondb,eckle2008attosecond,pfeiffer2012attoclock,pfeiffer2013recent,sainadh2019attosecond} while some attoclock experiments \cite{landsman2014ultrafast,camus2017experimental}, and a recent Larmor clock experiment \cite{ramos2020measurement} have reported non-zero tunneling time. 

The incorporation of time as a quantum observable remains controversial \cite{muga2007time}, however, one of us has shown that Pauli's no-go theorem does not hold within the single Hilbert space formulation of quantum mechanics and that Pauli made implicit assumptions that are inconsistent \cite{galapon2002pauli,galapon2002self,galapon2004confined,galapon2005confined}. Specifically, a self-adjoint time operator canonically conjugate to the Hamiltonian does exist, albeit in a closed subspace of the Hilbert space, i.e., there exists a non-zero vector orthogonal to the subspace. This means that a time operator is \textit{a priori} not self-adjoint unless its domain is specified, which opens the possibility of still considering time-of-arrival as a dynamical observable in quantum mechanics by requiring that a TOA-operator $\mathsf{\hat{T}}$ should, at the very least, be Hermitian \cite{galapon2009quantum,sombillo2014quantum,sombillo2016particle,galapon2018quantizations,flores2019quantum,pablico2020quantum}. 

The TOA-operator formalism uses the rigged Hilbert space (RHS) formulation of quantum mechanics to construct $\mathsf{\hat{T}}$ in coordinate representation. It was earlier applied by one of us to the tunneling time problem \cite{galapon2012only} by using the Weyl-ordering rule to quantize the classical TOA at the origin given by
\begin{subequations}
\begin{align}
\mathcal{T}_C(q,p) =& -\text{sgn}(p)\sqrt{\frac{\mu}{2}}\int_{0}^{q} \frac{dq'}{\sqrt{H(q,p)-V(q')}} \label{eq:classtoa} \\
H(q,p) =& \dfrac{p^2}{2\mu} + V(q)
\end{align}
\end{subequations}
where, $H(q,p)$ is the Hamiltonian, $V(q)$ is the interaction potential, and $\text{sgn}(z)$ is the signum function. The tunneling time is extracted from the difference of the expectation values of the TOA-operators in the presence and absence of a potential barrier. It was found that below-barrier components are transmitted instantaneously, i.e., tunneling time is instantaneous \cite{galapon2012only}. This result supports the predictions of Refs. \cite{torlina2015interpreting,petersen2017tunneling,petersen2018instantaneous}, which used a parametric approach to time, as well as the attoclock measurements done in Refs. \cite{eckle2008attosecondb,eckle2008attosecond,pfeiffer2012attoclock,pfeiffer2013recent,sainadh2019attosecond}. An alternative spacetime-symmetric extension of quantum mechanics has been recently proposed in Ref. \cite{dias2017space} (in this formalism time becomes a self-adjoint operator and position is a parameter) which was applied to the tunneling time problem by Ref. \cite{de2024traveling} and showed that instantantaneous tunneling time is still observed. Surprisingly, the same phenomenon also persists in a recently proposed formalism on the construction of relativistic TOA-operators via Weyl-ordering for spin-0 particles provided that the barrier height $V_o$ is less than the rest mass energy \cite{flores2022relativistic,flores2023instantaneous,flores2023quantized}. Additionally, it has been argued that only classically accessible trajectories contribute to the path integral \cite{jacak2023forbidden}, which implies that the tunneling time should be instantaneous since ``a Feynman path does not traverse the barrier region."  

Now, a fundamental problem with the approach done in Ref. \cite{galapon2012only} is that there are infinitely many possible TOA-operators\footnote{The superscript `Q' will be used to denote quantities related to the TOA-operators quantized via any non-Weyl ordering rule}, $\mathsf{\hat{T}^{Q}}$, corresponding to the same classical observable $\mathcal{T}_C(q,p)$. Canonical quantization of the classical TOA involves promoting the position and momentum observables $(q,p)$ into non-commutative operators $(\mathsf{\hat{q}},\mathsf{\hat{p}})$, which results to an ambiguity in the operator ordering, and since the resulting operators $\mathsf{\hat{T}^{Q}}$ are different, then the measurable observables will also be different. 

This ordering ambiguity raises the question of whether a preferred ordering rule exists (see Ref. \cite{carosso2022quantization} and references therein), however, it has been argued that these various ordering rules correspond to different experimental setups \cite{ali1990ordering,agarwal1970calculus}. For example, photoelectron correlation and coincidence experiments naturally express observables in terms of normally-ordered products of the annihilation and creation operators, while light scattering uses time-ordered operators \cite{agarwal1970calculus}. Thus, the results in Ref. \cite{galapon2012only} raise the following questions: First, is the predicted instantaneous tunneling time merely a consequence of using the Weyl-ordering rule? Second, do different ordering rules correspond to different parametric definitions of tunneling time such as Wigner phase time, Larmor time, etc.?

While canonical quantization of the classical TOA and the role of operator ordering rules seems promising to investigate the tunneling time problem, it does not exhaust all possible quantum images of the classical TOA. Pioneering work by Mackey on the construction of quantum observables based on first quantum principles showed that the position and momentum operators, and their commutation relation can be derived based solely on the assumption of homogeneity of free space \cite{mackey1968induced,mackey1978unitary,birkhoff1975logic}. This has also lead one of us to construct a TOA-operator $\mathsf{\hat{T}^{S}}$ without quantization using a method termed \textit{supraquantization} \cite{galapon2004shouldn}. 

The operator\footnote{The superscript `S' will be used to denote quantities related to the supraquantized TOA-operators} $\mathsf{\hat{T}^{S}}$ is constructed by imposing the time-energy canonical commutation relation (TE-CCR), Hermiticity, time-reversal symmetry, and invoking the quantum-classical correspondence principle. We wish to emphasize that contrary to $\mathsf{\hat{T}^{S}}$, the operators $\mathsf{\hat{T}^{Q}}$ generally do not satisfy the TE-CCR because it was never imposed in the first place. This now leads us to our third question: Will imposing the TE-CCR yield a non-zero tunneling time?

Here, we address all these questions and show that all the quantum images of the classical TOA, $\mathsf{\hat{T}^{Q}}$ and $\mathsf{\hat{T}^{S}}$, will yield an instantaneous tunneling time, i.e., neither the operator ordering rules nor TE-CCR play a role in the predicted instantaneous tunneling time. The rest of the paper is structured as follows: In Sec. \ref{sec:toaopr}, we provide a brief review of TOA operators and their construction within the RHS framework of quantum mechanics. We also present a unified approach on how to exhaust all possible Hermitian operator ordering rules on the construction of $\mathsf{\hat{T}^{Q}}$ by introducing a `deformation operator' that transforms the Weyl-ordered TOA-operator into any ordering rule. In Sec. \ref{sec:review}, we provide a review on the application of the theory of TOA-operators to the tunneling time problem, laying the foundation for investigating the role of operator ordering rules and the TE-CCR in Sec. \ref{sec:tunnelelingall}. Now, a self-consistent theory must yield the same result for coordinate and momentum-space representation which leads us to tackle the problem anew in Sec. \ref{sec:eigs} using the momentum-space representation of the TOA-operators and eigenfunctions. In Sec. \ref{sec:toadist} we demonstrate the full extent of the predicted instantaneous tunneling time and completeness of the TOA-eigenfunctions by comparing the TOA distributions of a free particle and a particle that tunneled through the barrier. Last, we conclude in Sec. \ref{sec:conc}.


\section{Theory of time-of-arrival operators}
\label{sec:toaopr}

\subsection{Operators in the rigged Hilbert space}

The postulates of quantum mechanics states that a quantum system is a Hilbert space $\mathcal{H}$ wherein the physical states are represented by unit rays $|\Psi\rangle$ in $\mathcal{H}$, while observable quantities are represented by linear self-adjoint operators $\mathsf{\hat{A}}$ acting on $\mathcal{H}$. The eigenvalues of these operators then represent the possible measurement outcomes of the corresponding observable and its spectrum may be discrete (particle in a box), continuous (free particle), or a combination of both (hydrogen atom). If $\mathsf{\hat{A}}$ is bounded and its spectrum is discrete, then $\mathsf{\hat{A}}$  is defined on $\mathcal{H}$ and its eigenvectors belong to $\mathcal{H}$. 

However, operators in quantum mechanics are generally unbounded with a continuous spectrum corresponding to non-normalizable eigenfunctions. For example, the Dirac-delta function and plane wave are eigenfunctions of the position and momentum observable, respectively, which are not square integrable and thus, do not belong to $\mathcal{H}$. In order to deal with these non-square integrable functions that are outside the Hilbert space, one can use Dirac’s bra-ket notation which is made mathematically rigorous by the rigged Hilbert space (RHS) through the theory of distributions \cite{de2005role}. The RHS is a triad of spaces $\Phi \subset \mathcal{H} \subset \Phi^\times$, where $\Phi$ is the space of test functions, and $\Phi^\times$ the space of distributions. The standard Hilbert space formulation of quantum mechanics is recovered by taking the closures on $\Phi$ with respect to the metric of $\mathcal{H}$. Here, we shall choose the fundamental space of our RHS to be the space of infinitely continuously differentiable complex valued functions with compact supports $\Phi$.

In coordinate representation, the TOA-operators have the general form 
\begin{align}
    (\mathsf{\hat{T}} \phi) (q) =&  \int_{-\infty}^\infty dq'  \langle q | \mathsf{\hat{T}}  | q' \rangle \phi(q') \nonumber \\
    =& \int_{-\infty}^\infty dq' \dfrac{\mu}{i\hbar} T(q,q') \text{sgn}(q-q') \phi(q'),
    \label{eq:rhstoa}
\end{align}
where, $\langle q | \mathsf{\hat{T}}  | q' \rangle = (\mu/i\hbar)T(q,q') \text{sgn}(q-q')$, and $T(q,q')$ is referred to as the time kernel factor (TKF). Eq. \eqref{eq:rhstoa} indicates that the TOA-observable $\mathsf{\hat{T}}$ is a mapping from $\Phi$ to $\Phi^\times$, and is interpreted as a functional on $\Phi$ wherein the kernel $\langle q | \mathsf{\hat{T}} | q' \rangle$ is  a distribution. Moreover, the TKF satisfies the following physical properties: 
\begin{enumerate}[(i)]
\item Hermiticity: $\mathsf{\hat{T}}=\mathsf{\hat{T}}^\dagger \rightarrow \langle q|\mathsf{\hat{T}}|q'\rangle = \langle q'|\mathsf{\hat{T}}|q\rangle^* \rightarrow T(q,q') = T(q',q)$, 
\item time-reversal symmetry: $\mathsf{\hat{\Theta}} \mathsf{\hat{T}} \mathsf{\hat{\Theta}^{-1}} = -\mathsf{\hat{T}} \rightarrow \langle q|\mathsf{\hat{T}}|q'\rangle^* = -\langle q|\mathsf{\hat{T}}|q'\rangle \rightarrow T(q,q')^* = T(q,q')$, and
\item  the quantum-classical correspondence principle established by the known Weyl-Wigner transform \cite{galapon2004shouldn,pablico2022quantum}
\begin{align}
\mathcal{T}_C(q,p) = \lim_{\hbar\rightarrow 0} \int_{-\infty}^\infty d\nu  \left\langle q + \frac{\nu}{2} \right| \mathsf{\hat{T}}  \left| q- \frac{\nu}{2} \right\rangle e^{-i\nu p/ \hbar}.
\end{align}
\end{enumerate}
The problem of constructing a TOA-operator thus translates to constructing the corresponding TKF.  

\subsection{Quantizing the classical time-of-arrival}
\label{subsec:quantizing}

It is easy to see that $\mathcal{T}_C(q,p)$, Eq. \eqref{eq:classtoa}, can be multiple and/or complex-valued. For example, a particle that is fired upward will cross the arrival point twice if the classical turning point is above the arrival point. Similarly, if the classical turning point is below the arrival point, then $\mathcal{T}_C(q,p)$ becomes complex-valued which indicates non-arrival. Thus, it has been deemed not meaningful to quantize $\mathcal{T}_C(q,p)$ \cite{leon2000time,peres1997quantum}. Ref. \cite{galapon2018quantizations} addressed objections against the quantization of $\mathcal{T}_C(q,p)$ based on physical arguments in the sense that:
\begin{enumerate}[(i)]
    \item the TOA of a quantum particle is always real-valued since it can tunnel through the classically forbidden region; and 
    \item the TOA of a quantum particle is always single-valued since a detector that registers the particle's arrival will result to a collapse of the wavefunction.  
\end{enumerate}
Therefore, it is only meaningful to quantize the first TOA. Different ordering rules will result to different TKFs $T(q,q')$ in Eq. \eqref{eq:rhstoa}, and Ref. \cite{galapon2018quantizations} was only able to perform the quantization using Weyl, Born-Jordan, and simple symmetric ordering rule as these are the most well-studied\footnote{Specifically, Weyl ordering preserves the covariant property of Hamiltonian dynamics with respect to linear canonical transforms \cite{Gosson2016,de2006symplectic}, Born-Jordan ordering preserves the equivalence of the the Schr\"{o}dinger and Heisenberg formulation of quantum mechanics \cite{Gosson2016,de2013born,cohen1966generalized}, and simple-symmetric ordering just provides the easiest possible ordering by using the ``average rule'' \cite{domingo2015generalized,shewell1959formation}.} ordering rules \cite{Gosson2016,de2006symplectic,de2013born,cohen1966generalized,domingo2015generalized,shewell1959formation,de2016born,cohen2012weyl,Gosson2016a,Gosson2011}.

The quantization of Eq. \eqref{eq:classtoa} is done by expanding $\mathcal{T}_C(q,p)$ around the free TOA such that it is single and real-valued \cite{galapon2018quantizations}, i.e., 
\begin{align}
\mathcal{T}_C(q,p) =& \sum_{m=0}^\infty  \dfrac{(2m-1)!!}{m! }\dfrac{(-\mu)^{m+1} }{p^{2m+1}}  \int_0^q dq' (V(q)-V(q'))^m,	\nonumber \\
=& \sum_{m=0}^\infty  \dfrac{(2m-1)!! (-\mu)^{m+1}}{m! }  \sum_{n=1}^\infty a_n^{(m)} q^n p^{-(2m+1)}. 
\label{eq:ltoa2}
\end{align}
The second line follows from the assumption that the potential $V(q)$ is analytic at the origin $q=0$ such that it admits the expansion $V(q)=\sum_{n=0}^\infty a_n q^n$. Notice that the above infinite series converges absolutely and uniformly to the classical arrival time only when the condition $|V(q)-V(q')|<p^2/2\mu$ is satisfied. Otherwise, the series diverges implying non-arrival at the arrival point. This characteristic of $\mathcal{T}_C(q,p)$ is particularly important in cases without classical counterparts, such as as the case of quantum tunneling, where the diverging expansion of $\mathcal{T}_C(q,p)$ reflects the non-arrival of a classical particle at the chosen arrival point.

The classical observables $(q,p)$ are then promoted into operators $(\mathsf{\hat{q}}, \mathsf{\hat{p}})$ such that the TOA-operator has the expansion 
\begin{align}
    \mathsf{\hat{T}^{Q}} =& \sum_{m=0}^\infty  \dfrac{(2m-1)!! (-\mu)^{m+1}}{m! }  \sum_{n=1}^\infty a_n^{(m)} \mathsf{\hat{t}^Q_{n,-2m-1}},
    \label{eq:toaexpansion}
\end{align}
where, $\mathsf{\hat{t}^Q_{n,-2m-1}}$ denotes a specific ordering on the quantization of the monomial $q^np^{-(2m+1)}$. Explicitly, the quantization is done by generalizing the Bender-Dunne basis operators \cite{bender1989exact,bender1989integration,bender2012matrix} given by 
\begin{equation}
	\mathsf{\hat{t}^Q_{n,-2m-1}} = \dfrac{\sum_{k=0}^n \beta_k^{(n)} \mathsf{\hat{q}}^k\mathsf{\hat{p}}^{-2m-1}\mathsf{\hat{q}}^{n-k}}{\sum_{k=0}^n \beta_k^{(n)}}, 
	\label{eq:BDbasis}
\end{equation} 
in which $\beta_k^{(n)}=\beta_{n-k}^{(n)*}$ to ensure that the ordering is Hermitian, e.g., 
\begin{align}
	\beta_k^{(n)} = 
	\begin{cases}
		\dfrac{n!}{k!(n-k)!} \quad &, \quad \text{Weyl} \\
		1 \quad &, \quad \text{Born-Jordan} \\
		\delta_{k,0} + \delta_{k,n} \quad &, \quad \text{simple-symmetric}.
	\end{cases}
	\label{eq:coeff}
\end{align}
However, this now opens the problem on how to exhaust all possible Hermitian ordering rules and obtain the corresponding TKF $T^Q(q,q')$.

\begin{figure}[t!]
\centering
\includegraphics[width=0.45\textwidth]{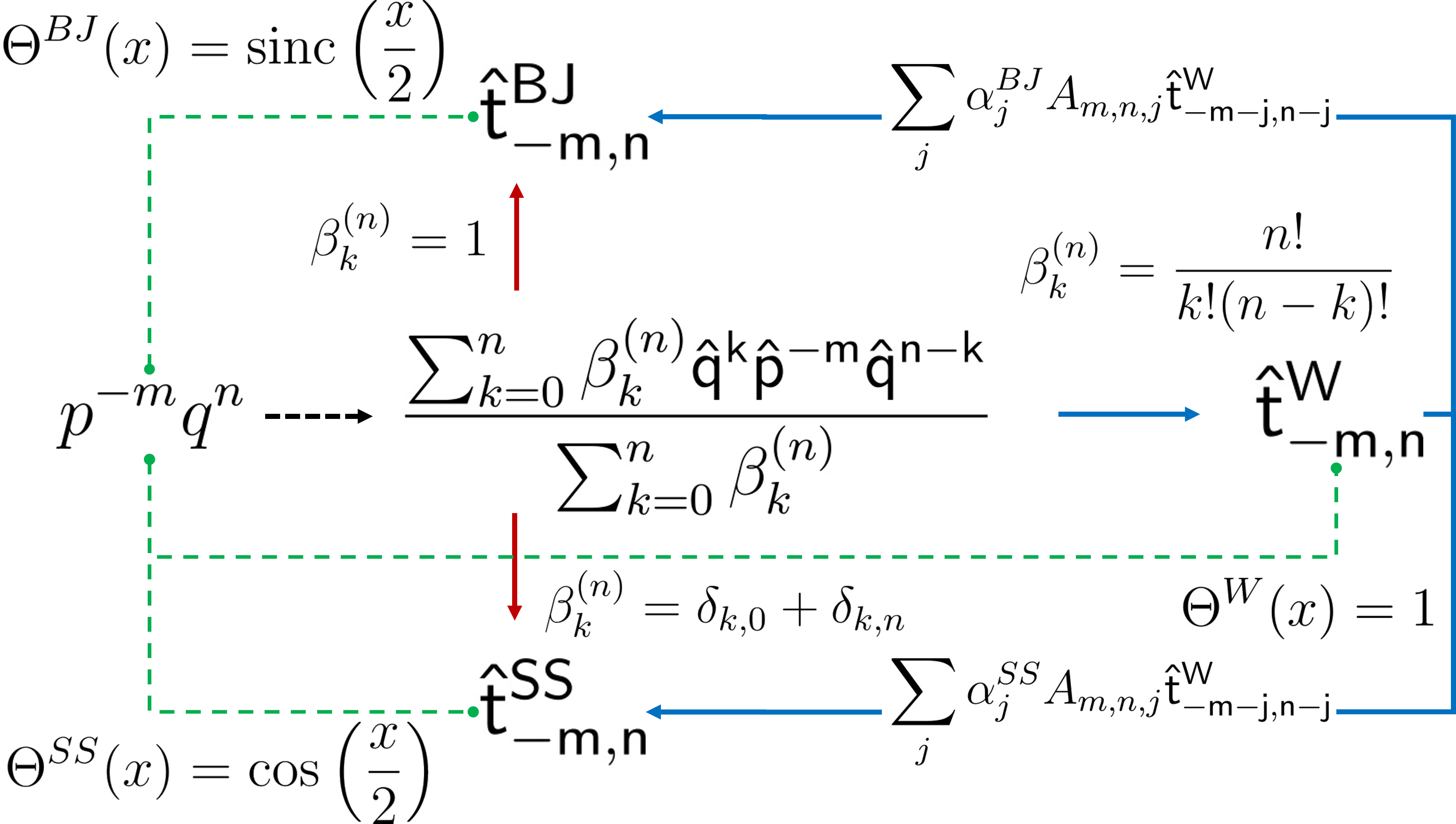}
\caption{Visual representation of obtaining the operator $\mathsf{\hat{T}^Q}$ through various ordering rules. Red arrows indicate the `direct route' while the blue arrows indicate the `indirect route' via the ordering function $\Theta(x)$ to facilitate the transformation from Weyl to all other ordering rules. }
\label{fig:transform}
\end{figure}

It was shown in Ref. \cite{domingo2015generalized} that every known ordering rule can be associated to a real-valued function $\Theta^Q(x)= \sum \alpha_j^Q x^j$, where $\Theta^Q(0)=1$. The ordering function $\Theta^Q(x)$ then gives rise to transformation equations from Weyl to all other ordering rules\footnote{For example, the ordering function for the Weyl, Born-Jordan and simple symmetric ordering are $\Theta^W(x)=1$, $\Theta^{BJ}(x)=\text{sinc}(x/2)$, and $\Theta^{SS}(x)=\cos(x/2)$, respectively \cite{domingo2015generalized}.}. The results of Ref. \cite{domingo2015generalized} now implies that instead of exhausting all possible $\beta_k^{(n)}$, we can simply use the Weyl-ordered Bender-Dunne basis operators and use the ordering function $\Theta^Q(x)$ to facilitate the transformation\footnote{In general, the correspondence between the coefficients $\beta_k^{(n)}$ in the Bender-Dunne basis operators with a specific ordering function is unknown, and Ref. \cite{domingo2015generalized} only lists the known ordering functions.} of the Weyl-ordered TKF $T^W(q,q')$ into all other ordering rules $T^Q(q,q')$, see Fig. \ref{fig:transform}. For the case of $\mathsf{\hat{t}^Q_{n,-2m-1}}$, the transformation is given by \cite{domingothesis},
\begin{subequations}
\begin{align}
\mathsf{\hat{t}^Q_{n,-2m-1}} =& \sum_{j=0}^\infty \dfrac{\hbar^j \alpha_j^Q  \Gamma(-2m)\Gamma(n+1)}{\Gamma(-2m-j)\Gamma(n-j+1)} \mathsf{\hat{t}^{W}_{n-j,-2m-1-j}}, \label{eq:allordering} \\
\mathsf{\hat{t}^{W}_{N,-M}} =& \dfrac{1}{2^{N+1}} \sum_{k=0}^N \dfrac{N!}{(N-k)!k!} \mathsf{\hat{q}^k} \mathsf{\hat{p}^{-M}} \mathsf{\hat{q}^{N-k}}. \label{eq:weyl}
\end{align}
\end{subequations}

Using Eqs. \eqref{eq:toaexpansion}, \eqref{eq:allordering}, and \eqref{eq:weyl}, the coordinate representation of any non-Weyl ordered TOA-operator now admits the expansion
\begin{align}
(\mathsf{\hat{T}^Q}\phi)(q) =& \int_{-\infty}^\infty dq' \langle q | \mathsf{\hat{T}^Q} | q' \rangle \phi(q'), \nonumber \\
=&  \dfrac{1}{2} \int_{-\infty}^\infty dq' \phi(q') \text{sgn}(q-q') \sum_{m=0}^\infty \dfrac{(2m-1)!!}{(2m)!m!}  \nonumber \\
&\times \left( \dfrac{\mu}{\hbar^2} (q-q')^2 \right)^m  \sum_{n=1}^\infty a_n^{(m)}  \sum_{j=0}^\infty \alpha_j^Q  (-i)^j \nonumber \\
&\times  (q-q')^j \dfrac{\Gamma(n+1)}{\Gamma(n-j+1)} \left(\dfrac{q+q'}{2}\right)^{n-j},
\label{eq:toaexpansion2}
\end{align}
wherein, we used the identity
\begin{align}
    \langle q | \mathsf{\hat{p}^{-k}} | q' \rangle =& \dfrac{i^k}{2\hbar^k(k-1)!}(q-q')^{k-1}\text{sgn}(q-q'),
    \label{eq:momentumkernel}
\end{align}
for $k \in \mathbb{Z}^+$ \cite{gelgeneralized}. A closed form expression of Eq. \eqref{eq:toaexpansion2} can be obtained by performing a change of variables $\eta=(q+q')/2$ and $\zeta=q-q'$ so that we can rewrite 
\begin{align}
	\dfrac{\Gamma(n+1)}{\Gamma(n-j+1)} \left(\dfrac{q+q'}{2}\right)^{n-j} = \dfrac{d^j \eta^n}{d\eta^j}.
\end{align}
Performing the necessary calculations, we then obtain
\begin{subequations}
\begin{align}
    (\mathsf{\hat{T}^Q}\phi)(q) =& \int_{-\infty}^\infty dq' \dfrac{\mu}{i\hbar} \bigg[ \mathsf{\hat{\Theta}^Q}  \tilde{T}^W(\eta,\zeta) \bigg] \text{sgn}(q-q') \phi(q') \label{eq:toaRHS},\\
    \tilde{T}^W(\eta,\zeta) =&  \dfrac{1}{2} \int_0^\eta ds \, {_0}F_1\left[ ; 1; \dfrac{\mu}{2 \hbar^2}(V(\eta)-V(s))\zeta^2\right], \label{eq:weyltkf}\\
    \mathsf{\hat{\Theta}^Q} =& \Theta^Q \left( -i\zeta\frac{d}{d\eta} \right)= \sum_{j=0}^\infty \alpha_j^Q (-i\zeta)^j \dfrac{d^j}{d\eta^j}, \label{eq:deformation} 
\end{align}    
\end{subequations}
in which, $T(q,q')=\tilde{T}(\eta,\zeta)$, and ${_0}F_1(;1;z)$ is a specific hypergeometric function. See Appendix \ref{app:deformation} for details.  

The ordering function $\Theta^Q(x)$ now serves as a `deformation operator', $\mathsf{\hat{\Theta}^Q}$, that acts on $T^{W}(q,q')$ to obtain the corresponding TKF associated to all other ordering rules. Moreover, in the $(\eta,\zeta)$-coordinates, Hermiticity implies that the TKF $\tilde{T}^Q(\eta,\zeta)=\mathsf{\hat{\Theta}^Q}  \tilde{T}^W(\eta,\zeta)$ satisfies $\tilde{T}^Q(\eta,\zeta)=\tilde{T}^Q(\eta,-\zeta)$. This means that the ordering function $\Theta^Q(x)$ is always an even function since the Weyl-ordered TKF is already Hermitian, i.e., $\tilde{T}^W(\eta,\zeta)=\tilde{T}^W(\eta,-\zeta)$. Equations \eqref{eq:toaRHS}-\eqref{eq:deformation} now presents a unified description on the construction of all possible Hermitian ordering rules on the canonical quantization of the classical TOA. Appendix \ref{app:examples} demonstrates the validity of the TKF transformation Eqs. \eqref{eq:toaRHS}-\eqref{eq:deformation} using the linear and harmonic oscillator potential as examples by recovering the Born-Jordan and simple-symmetric ordered TKFs obtained in Ref. \cite{galapon2018quantizations}.

\subsection{Supraquantized TOA-operators}\label{subsec:supraquantization}

While canonical quantization appears to be a promising solution to the quantum TOA problem, a fundamental issue on the operators $\mathsf{\hat{T}^Q}$ is that they generally do not satisfy the TE-CCR as it was never imposed in the first place. Second, there exists obstructions to quantization \cite{groenewold1946principles,van1951certaines,gotay1999groenewold}, i.e. no quantization exists such that for all classical observables $f$ and $g$ the Dirac condition is satisfied, i.e., $\{f,g\}  \rightarrow [ \mathsf{\hat{Q}_F} , \mathsf{\hat{Q}_g} ] = i\hbar \mathsf{\hat{Q}_{ \{f,g\} }}$
where, $\{f,g\}$ denotes the Poisson bracket. While all operator ordering rules lead to the same conjugacy-preserving TOA operator for the free-particle case\footnote{See Eq. \eqref{eq:BDbasis} for the monomial $qp^{-1}$}, they are completely different for the interacting case. Only the Weyl-ordered TOA-operator $\mathsf{\hat{T}^W}$ satisfies the TE-CCR but is limited to linear systems, i.e., $V(q)=aq^2+bq+c$ \cite{galapon2004shouldn}. For nonlinear systems, a solution does not necessarily exist due to obstructions to quantization \cite{domingothesis,groenewold1946principles,van1951certaines,gotay1999groenewold}. 

Inspired by the works of Mackey \cite{mackey1968induced,mackey1978unitary,birkhoff1975logic}, one of us has reconsidered the quantum TOA problem by constructing a TOA operator $\mathsf{\hat{T}^S}$ independent of canonical quantization \cite{galapon2004shouldn}. The method was called \textit{supraquantization} where quantum observables are constructed from first principles, and starts by imposing the TE-CCR, wherein, for every $\phi$ and $\varphi$ in $\Phi$, the TOA-operator $\mathsf{\hat{T}^S}$ must satisfy
\begin{align}
\langle \phi | [ \mathsf{\hat{H}} , \mathsf{\hat{T}^S} ] | \varphi \rangle = i\hbar \langle \phi | \varphi \rangle . 
\label{eq:teccr}
\end{align}
The commutator on the left-hand side of Eq. \eqref{eq:teccr} is then expanded, and the identity $\int dq |q\rangle \langle q |$ is inserted to the resulting expression, i.e., 
\begin{align}
\int_{-\infty}^\infty dq \int_{-\infty}^\infty & dq' \langle \phi | \mathsf{\hat{H}} | q \rangle \langle q| \mathsf{\hat{T}^S}  | q' \rangle \langle q' | \varphi \rangle \nonumber \\
- \int_{-\infty}^\infty & dq \int_{-\infty}^\infty dq' \langle \phi | q \rangle \langle q| \mathsf{\hat{T}^S}  | q' \rangle \langle q' | \mathsf{\hat{H}} | \varphi \rangle \nonumber \\
&= i\hbar \int_{-\infty}^\infty dq \phi^*(q) \varphi(q),
\end{align}
where the kernels are given by 
\begin{align}
\langle q| \mathsf{\hat{T}^S}  | q' \rangle =& \dfrac{\mu}{i\hbar} T^S(q,q') \text{sgn}(q-q') \\
\langle q' | \mathsf{\hat{H}} | \varphi \rangle =& -\dfrac{\hbar^2}{2\mu}\dfrac{d^2 \varphi(q')}{dq'^2} + V(q') \varphi(q') \\
\langle \phi | \mathsf{\hat{H}} | q \rangle =& -\dfrac{\hbar^2}{2\mu}\dfrac{d^2 \phi^*(q)}{dq^2} + V(q) \phi^*(q). 
\end{align}
Performing the necessary operations and imposing the relevant physical properties similar to $\mathsf{\hat{T}^S}$, i.e., Hermiticity, time-reversal symmetry, and quantum-classical correspondence, leads to a second order partial differential equation  
\begin{align}\label{TKEorigintro}
-\frac{\hbar^2}{2\mu}  \frac{\partial^2 T^S(q,q')}{\partial q^2}+&\frac{\hbar^2}{2\mu} \frac{\partial^2 T^S(q,q')}{\partial q'^2} \nonumber \\
&+ \left[V(q)-V(q')\right]T^S(q,q')=0,
\end{align}
for the TKF $T^S(q,q')$ which is referred to as the time kernel equation, and subject to the boundary conditions
\begin{subequations}
\begin{align}
T^S(q,q) =& \dfrac{q}{2},  \\
T^S(q,-q) =& 0.
\end{align}
\end{subequations}
The full solution of Eq. \eqref{TKEorigintro} has been recently obtained in Ref. \cite{pablico2022quantum} and is given by
\begin{subequations}
\begin{equation}
\tilde{T}^S(\eta,\zeta) =\tilde{T}^W(\eta,\zeta) +  \sum_{n=1}^{\infty}\tilde{T}^{S,(n)}(\eta,\zeta),
\label{tkesol3compact}
\end{equation}
\begin{align}
\tilde{T}^{S,(n)}(\eta,\zeta) =& \left(\frac{\mu}{2\hbar^2}\right)\sum_{r=1}^{n} \dfrac{1}{4^{2r}(2r+1)!} \int_0^\eta ds \dfrac{d^{2r+1}V(s)}{ds^{2r+1}} \nonumber \\
&\times \int_0^\zeta dw w^{2r+1} G(s,w) \tilde{T}^{S,(n-r)}(s,w), \label{tnuvintro}
\end{align}
\begin{align}
G(s,w) =&\, {_0}F_1 \left[ ; 1 ; \left(\dfrac{\mu}{2\hbar^2}\right)(\zeta^2-w^2) \{ V(\eta) - V(s) \} \right],
\label{gsw}
\end{align}
\end{subequations}
wherein the zeroth order term is exactly the Weyl-ordered TKF, i.e., $\tilde{T}^{S,(0)}(\eta,\zeta)= \tilde{T}^W(\eta,\zeta)$. The terms $\tilde{T}^{S,(n)}(\eta,\zeta)$ are interpreted as the quantum corrections to $\tilde{T}^W(\eta,\zeta)$ which are needed to satisfy the conjugacy relation, and  vanish for linear systems.

\section{Review of the TOA-operators and tunneling time}
\label{sec:review}

\begin{figure}[t!]
\centering
\includegraphics[width=0.45\textwidth]{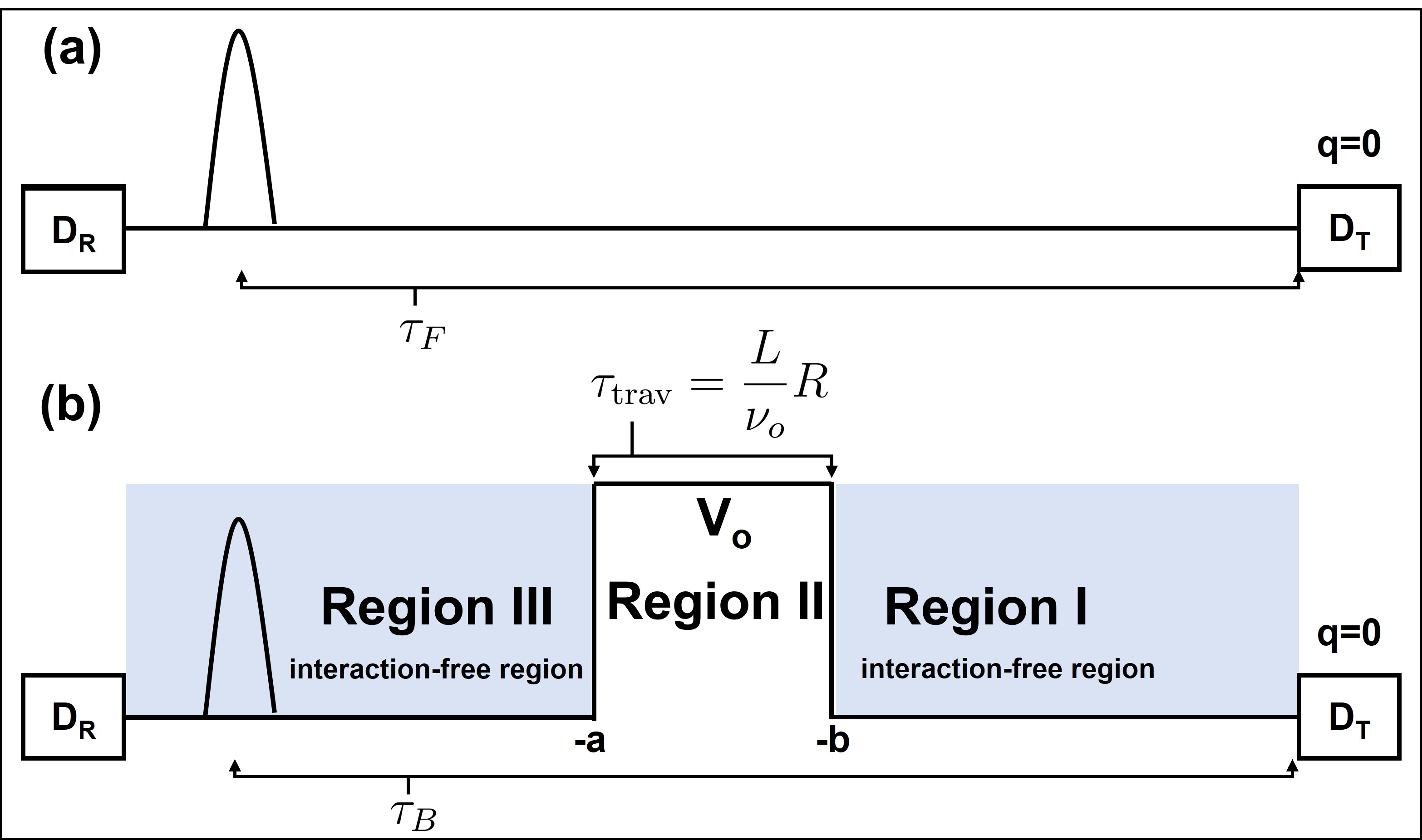}
\caption{Measurement scheme for the expected quantum traversal time. The detector $D_T$ is placed at the arrival point $q=0$ and records the arrival of the particle, while the detector $D_R$ is placed at the far left of the barrier and does not record the particle's arrival.}
\label{fig:setup}
\end{figure}

For completeness, we provide a discussion on how the theory of TOA-operators was applied to the tunneling time problem \cite{galapon2012only}. The method will then serve as foundation for the calculations in Sec. \ref{sec:tunnelelingall}. The analysis starts by considering an incident wavepacket $\psi(q)=e^{ik_o q}\varphi(q)$ that is initially centered at $q=q_o$ with average momentum $p_o=\hbar k_o$. A square barrier is then placed between $q_o$ and the arrival point $q=0$, i.e., $V(q)=V_o$ for $-a<q<-b$. Following the measurement scheme in Fig. \ref{fig:setup}, the average arrival time is then assumed to be the expectation value of $\mathsf{\hat{T}_B^W}$, i.e., $\bar{\tau}_B^W=\langle\psi|\mathsf{\hat{T}_B}^W|\psi\rangle$. The same scheme is then applied in the absence of a barrier with a corresponding TOA-operator $\mathsf{\hat{T}_F^W}$. The subscripts `F' and `B' indicate quantities related to the absence and presence of the barrier, respectively. The tunneling time is then extracted from the difference of the average TOA in the presence and absence of the barrier, that is, 
\begin{subequations}
\begin{equation}
    \Delta\bar{\tau}^W=\bar{\tau}_F^W-\bar{\tau}_B^W=\langle \psi | \mathsf{\hat{T}_F^W} | \psi \rangle - \langle \psi | \mathsf{\hat{T}_B^W} | \psi \rangle,
    \label{eq:toadiff0}
\end{equation} 
\begin{align}
    \bar{\tau} =& \dfrac{\mu}{i\hbar} \int_{-\infty}^\infty dq \int_{-\infty}^\infty dq' \psi^*(q) T(q,q') \text{sgn}(q-q') \psi(q').
\end{align}
\end{subequations}

In the absence of the barrier, the Weyl-ordered TKF $\tilde{T}_F^W(\eta,\zeta)$ is constructed by substituting $V(q)=0$ into Eq. \eqref{eq:weyltkf} which yields
\begin{equation}
    \tilde{T}_{F}^{W}(\eta) = \frac{\eta}{2},
    \label{eq:weylfreetkf}
\end{equation}
and coincides with the RHS extension of the well-known Aharonov-Bohm \cite{aharonov1961time} free TOA-operator 
\begin{equation}
    \mathsf{\hat{T}^{AB}}=-\dfrac{\mu}{2}(\mathsf{\hat{q}}\mathsf{\hat{p}^{-1}}+\mathsf{\hat{p}^{-1}}\mathsf{\hat{q}}).
\end{equation}
In the presence of the barrier, the Weyl-ordered TKF $\tilde{T}_B^W(\eta,\zeta)$ is constructed by mapping the potential $V(q)$ into the $\eta$-coordinate such that $V(\eta)=V_o$ for $-a<\eta<-b$ and $V(\eta)=0$ outside this interval. The integral Eq. \eqref{eq:weyltkf} is then split into three non-overlapping regions separated by the edges of the barrier. The TKF $\tilde{T}_B^W(\eta,\zeta)$ thus have three pieces which depends on the support of the incident wavepacket $\psi(q)$ as shown in Fig. \ref{fig:setup}, i.e.,    
\begin{subequations}
\begin{align}
\tilde{T}_{B,I}^W(\eta,\zeta)=&\frac{\eta}{2}, \label{eq:barrierTKFweyl1} \\
\tilde{T}_{B,II}^W(\eta,\zeta)=&\frac{\eta+b}{2}-\frac{b}{2}I_0(\kappa_o |\zeta|), \label{eq:barrierTKFweyl2} \\
\tilde{T}_{B,III}^W(\eta,\zeta)=&\frac{\eta+L}{2}-\frac{L}{2}J_0(\kappa_o |\zeta|), \label{eq:barrierTKFweyl3}
\end{align}
\end{subequations}
where $L=a-b$ is the length of the barrier, $\kappa_o=\sqrt{2\mu V_o}/\hbar$ while $I_0(z)$ and $J_0(z)$ are specific Bessel functions. It is assumed that the incident wavepacket $\psi(q)=e^{ik_oq}\varphi(q)$ only has support on the left-side of the barrier such that it does not initially `leak' into the barrier system and there is a zero probability that it is already at the transmission side at the initial time $t=0$. This means that only the piece $\tilde{T}_{B,III}^W(\eta,\zeta)$ is relevant for the barrier TOA-operator $\mathsf{\hat{T}_B^W}$. 

In the $(\eta,\zeta)$-coordinates, we can rewrite the TOA difference Eq. \eqref{eq:toadiff0} as
\begin{align}
\Delta\bar{\tau}^W = \frac{-2\mu}{\hbar}\mathrm{Im} & \left[ \int_{0}^{\infty} d\zeta \int_{-\infty}^{\infty} d\eta \, e^{i k _o\zeta} \tilde{T}_{F-B}^W(\eta,\zeta) \right. \nonumber \\
&\times   \left. \varphi^*\left(\eta-\frac{\zeta}{2}\right)\varphi\left(\eta+\frac{\zeta}{2}\right) \right] \label{eq:toadiff0}
\end{align}
where, $\text{Im}[.]$ denotes the imaginary component of the integral, and $\tilde{T}_{F-B}^W(\eta,\zeta)=\tilde{T}_{F}^{W}(\eta) - \tilde{T}_{B,III}^W(\eta,\zeta)$. Explicitly, Eq. \eqref{eq:toadiff0} simplifies to
\begin{subequations}
\begin{align}
\Delta\bar{\tau}^W =& \dfrac{\mu L}{\hbar} \mathrm{Im} \left\{ \int_{0}^{\infty} d\zeta \, e^{i k _o\zeta} \left[ 1 - J_0(\kappa_o |\zeta|) \right] \Phi(\zeta) \right\}, \\
\Phi(\zeta) =& \int_{-\infty}^{\infty} d\eta   \varphi^*\left(\eta-\frac{\zeta}{2}\right)\varphi\left(\eta+\frac{\zeta}{2}\right).
\end{align}
\end{subequations}
We then introduce the Fourier transform 
\begin{align}
\varphi(q) = \int_{-\infty}^\infty \dfrac{d\tilde{k}}{\sqrt{2\pi}} e^{i \tilde{k} q} \phi(\tilde{k}),
\end{align}
to investigate which components of the incident wavepacket $\psi(q)=e^{ik_oq}\varphi(q)$ tunnel through the barrier and contribute to the tunneling time. Performing the necessary operations, the TOA difference now takes the form 
\begin{subequations}
\begin{align}
\Delta\bar{\tau}^W=&\dfrac{L}{\nu_0} (Q^W-R^W) \label{eq:toadiff2}, \\
Q^W =&  \text{P.V.}\left[ k_o \int_{-\infty}^\infty dk \dfrac{|\tilde{\psi}(k)|^2}{k} \right], \label{eq:Q}\\
R^W =& k_o\int_{\kappa_o}^\infty dk \dfrac{|\tilde{\psi}(k)|^2-|\tilde{\psi}(- k)|^2}{\sqrt{k^2-\kappa_o^2}}, \label{eq:R}
\end{align}
\end{subequations}
where $\nu_o=p_o/\mu$ is the group velocity of the incident wavepacket, $\tilde{\psi}(k)$ is the Fourier transform of the incident wavepacket, and P.V. denotes the principal value. Here, we also used the integral identity 
\begin{align}
\int_0^\infty dx \, J_0(ax) \sin(bx) = \dfrac{H(b-a)}{\sqrt{b^2-a^2}},
\label{eq:integral}
\end{align}
where $H(z)$ is the Heaviside function \cite{gradshteyn2014table}. 

The physical significance of the quantities $Q^W$ and $R^W$ are obtained by considering the high energy limit $k_o\rightarrow\infty$. Specifically, this yields $Q^W\sim 1$ which physically suggest that the quantity $\tau_Q^W=(L/\nu_o)Q^W$ is just the free classical TOA across the region of length $L$. Similarly, $R^W \sim \nu_o/\nu$ where $\nu$ is the particle's speed on top of the barrier. This physically suggest that $R^W$ is an effective index of refraction of the barrier with respect to the incident wavepacket.

We interpret the negative sign in front of $|\tilde{\psi}(-k)|^2$ as a physical indication of the quantum particle's non-arrival at the transmission channel. Hence, for arrivals in the transmission channel, only the contribution of $\tilde{\psi}(+k)$ is physically relevant, and we can extract the barrier traversal time from Eq. \eqref{eq:toadiff2} as:
\begin{equation}
\tau_{\text{trav}}^W=\dfrac{L}{\nu_o}R^W = \dfrac{L}{\nu_o} \left[ k_o\int_{\kappa_o}^\infty dk \dfrac{|\tilde{\psi}(k)|^2}{\sqrt{k^2-\kappa_o^2}} \right]
\end{equation}
The vanishing of the momentum contributions $0<k<\kappa_o$ to $\tau_{\text{trav}}^W$ leads to the conclusion that tunneling time is instantaneous. Within the assumptions of the formalism, this can be realized by preparing a spatially broad wavepacket such that the spread in momentum is narrow to ensure that all the momentum components will be positive. Moreover, the incident wavepacket must then be initially placed sufficiently far from the barrier so that it does not `leak' into the barrier region at $t=0$.

\section{Tunneling time for all other TOA-operators}
\label{sec:tunnelelingall}

\subsection{Tunneling time for the operators $\mathsf{\hat{T}^Q}$}
\label{sec:tunnelingQ}

In Sec. \ref{sec:toaopr} we have shown how the Weyl-ordered TKF $T^W(q,q')$ can be `deformed' to obtain the corresponding TKF for all other ordering rules. Thus, it follows from Eqs. \eqref{eq:toaRHS}-\eqref{eq:deformation} that the TOA difference $\Delta\bar{\tau}^Q$ for all other ordering rules can be written as  
\begin{align}
\Delta\bar{\tau}^Q = \frac{-2\mu}{\hbar}\mathrm{Im} & \left[ \int_{0}^{\infty} d\zeta \int_{-\infty}^{\infty} d\eta \, e^{i k _o\zeta} \tilde{T}_{F-B}^Q(\eta,\zeta) \right. \nonumber \\
&\times   \left. \varphi^*\left(\eta-\frac{\zeta}{2}\right)\varphi\left(\eta+\frac{\zeta}{2}\right) \right] \label{eq:tkfdiffnonWeyl}
\end{align}
where
\begin{align}
\tilde{T}_{F-B}^Q(\eta,\zeta) = \mathsf{\hat{\Theta}^Q}\tilde{T}_{F}^{W}(\eta) - \mathsf{\hat{\Theta}^Q}\tilde{T}_{B,III}^W(\eta,\zeta),
\end{align}
It easily follows from Eq. \eqref{eq:weylfreetkf} that $\mathsf{\hat{\Theta}^Q}\tilde{T}_{F}^{W}(\eta) = \tilde{T}_{F}^{W}(\eta)$ since the `deformation operator' $\mathsf{\hat{\Theta}^Q}$ is an even-function and therefore, the only-non-zero term after applying $\mathsf{\hat{\Theta}^Q}$, Eq. \eqref{eq:deformation}, on $\tilde{T}_{F}^{W}(\eta)$ is the leading term $j=0$. Similar arguments follow for $\mathsf{\hat{\Theta}^Q}\tilde{T}_{B,III}^W(\eta,\zeta)=\tilde{T}_{B,III}^W(\eta,\zeta)$. Thus, Eq. \eqref{eq:tkfdiffnonWeyl} indicates that TOA difference is independent of the ordering rule and the results of Ref. \cite{galapon2012only} still holds, i.e., tunneling time is still instantaneous,  
\begin{align}
\Delta\bar{\tau}^Q=\Delta\bar{\tau}^W.
\end{align}
We emphasize that Eq. \eqref{eq:tkfdiffnonWeyl} is only valid because of the assumption that the incident wavepacket only has support on the left-side of the barrier. If the support of $\psi(q)$ extends to other regions, then the TKF $\tilde{T}^W(\eta,\zeta)$ will have jump discontinuities along the edges of the barrier and must be expressed as 
\begin{align}
    \tilde{T}_B^W(\eta,\zeta) =& H(-\eta-a)\tilde{T}_{B,III}^W(\eta,\zeta) \nonumber \\
    &+ \left\{ H(\eta+a) + H(-\eta-b) \right\} \tilde{T}_{B,II}^W(\eta,\zeta)  \nonumber \\
    &+ H(\eta+b)\tilde{T}_{B,I}^W(\eta,\zeta). 
\end{align}
Since $\tilde{T}^W(\eta,\zeta)$ are interpreted in a distributional sense, then applying the deformation operator $\mathsf{\hat{\Theta}^Q}$ might possibly introduce new terms that depends on the ordering rule.    

\subsection{Tunneling time for the operator $\mathsf{\hat{T}^S}$}

Let us now consider the quantum tunneling time problem when the TE-CCR is imposed. This suggests that the operator $\mathsf{\hat{T}_B} $ appearing in Eq. (\ref{eq:toadiff0}) is to be constructed by supraquantization (see Sec. \ref{subsec:supraquantization}). Since the leading term of the supraquantized operator is simply the Weyl-quantized TOA operator, it follows from Eqs. \eqref{tkesol3compact}-\eqref{gsw} that the TOA difference $\Delta\bar{\tau}^S$ appears as the following expansion
\begin{subequations}
\begin{align}
\Delta\bar{\tau}^S =& \Delta\bar{\tau}^W - \sum_{n=1} \bar{\tau}^{S,(n)}_{B}, \label{toadiff2expand}
\end{align}
\begin{align}
\bar{\tau}^{S,(n)}_{B}= \frac{-2\mu}{\hbar}\mathrm{Im} & \left[ \int_{0}^{\infty} d\zeta \int_{-\infty}^{\infty} d\eta \, e^{i k _o\zeta} \tilde{T}_{B}^{S,(n)}(\eta,\zeta)  \right. \nonumber \\
&\times   \left. \varphi^*\left(\eta-\frac{\zeta}{2}\right)\varphi\left(\eta+\frac{\zeta}{2}\right) \right] 
\end{align}
\end{subequations}
The full contribution of the TE-CCR is obtained by investigating the additional terms,  $\bar{\tau}^{S,(n)}_{B}$ for $n\geq1$, in Eq. (\ref{toadiff2expand}). 

Let us consider the leading TKF correction, the $n=1$ case, given by 
\begin{align}
\tilde{T}_{B}^{S,(1)}(\eta,\zeta) =  \frac{\mu}{24\hbar^2} & \int_{0}^{\eta} ds \dfrac{d^3 V(s)}{ds^3} \nonumber \\
&\times \int_{0}^{\zeta} dw \, w^3 G(s,w)  \tilde{T}_{B}^{W}(s,w). 
 \label{t10uvfinbbar}
\end{align}
Following the methods in Sec. \ref{sec:review}, we also need to split our integral along $s$ into the same three non-overlapping regions defined by regions $I$, $II$, and $III$. This means that the kernel factor $\tilde{T}_{B}^{S,(1)}(\eta,\zeta)$ has also three pieces corresponding to the possible locations of $\eta$. We interpret Eq. \eqref{t10uvfinbbar} in the distributional sense so that one has to be careful on dividing the integral and obtaining the partial derivatives of the interaction potential. To obtain the possible effects, if any, of the discontinuities at the edges of the potential barrier, we rewrite the potential barrier as
\begin{equation}\label{potinheav}
V(\eta)=V_0\,\big[H(\eta+a)-H(\eta+b)\big], 
\end{equation}
in $\eta$ coordinate where $H(x)$ is the usual Heaviside step function. Hence, the partial derivative involving the interaction potential is expressed in terms of derivatives of the Dirac delta function, that is,
\begin{equation}\label{potinheavd3}
V^{(2r+1)}(s)=2V_0\,\big[\delta^{(2r)}(s+a)-\delta^{(2r)}(s+b)\big],
\end{equation}
for $r \ge 0.$ 

For Region I defined by $\eta \geq -b$, we find the first piece given by
\begin{align}
\tilde{T}_{B,I}^{S,(1)}(\eta,\zeta)=&\frac{\mu V_0}{12\hbar^2}\int_{0}^{\eta} ds \, \big[\delta^{(2)}(s+a)-\delta^{(2)}(s+b)\big] \nonumber \\
&\times \int_{0}^{\zeta} dw \, w^3 \, \tilde{T}^{W}_{B,I}(s,w), \label{t10uvfinbbar11}
\end{align}
where $\tilde{T}^{W}_{B,I}(s,w)$ is the first piece of the Weyl-quantized barrier TKF. Substitution of Eq. (\ref{eq:barrierTKFweyl1}) into Eq. (\ref{t10uvfinbbar11}) and performing integration by parts twice yield
\begin{align}
\tilde{T}_{B,I}^{S,(1)}(\eta,\zeta)=&\frac{\mu V_0}{12\hbar^2}\int_{0}^{\eta} ds \, \big[\delta(s+a)-\delta(s+b)\big] \nonumber \\
&\times \frac{\partial^2}{\partial s^2}\left[\frac{s}{2}\,\int_{0}^{\zeta} dw \, w^3\right]\,=\,0. \label{t10uvfinbbar112}
\end{align}
The vanishing of $\tilde{T}_{B,I}^{S,(1)}$ follows from the fact that the factor in square brackets is only linear in $s$.

For Region II defined by $-a \le \eta \le -b$, we have to divide the integration along $s$ in Eq. (\ref{t10uvfinbbar}) into two parts corresponding to the regions $-b<s<0$ and $\eta<s<-b$. Hence, we have
\begin{align}
\tilde{T}_{B,II}^{S,(1)}(\eta,\zeta)=\frac{\mu V_0}{12\hbar^2} & \int_{0}^{-b} ds \, \big[\delta^{(2)}(s+a)-\delta^{(2)}(s+b)\big] \nonumber \\
&\times \int_{0}^{\zeta} dw \, w^3 \, \tilde{T}^{W}_{B,I}(s,w) \nonumber \\
+\frac{\mu V_0}{12\hbar^2} &\int_{-b}^{\eta} ds \,\big[\delta^{(2)}(s+a)-\delta^{(2)}(s+b)\big] \nonumber \\
&\times \int_{0}^{\zeta} dw \, w^3 \, \tilde{T}^{W}_{B,II}(s,w). \label{t10uvfinbbar12}
\end{align}
Substitution of Eqs. \eqref{eq:barrierTKFweyl1}-\eqref{eq:barrierTKFweyl2} and performing integration by parts twice yield 
\begin{equation}\label{t10uvfinbbar122}
\begin{split}
\tilde{T}_{B,II}^{S,(1)}(\eta,\zeta)=0,
\end{split}
\end{equation}
which again follows from the fact that the pieces $\tilde{T}^{W}_{B,I}$ and $\tilde{T}^{W}_{B,II}$ are only linear in $s$.

Last, for Region III defined by $\eta\leq-a$ we divide the integration along $s$ into three parts corresponding to the regions $-b<s<0$, $-a\le s\le b$, and $s<-a$, which yields
\begin{align}
\tilde{T}_{B,III}^{S,(1)}(\eta,\zeta)=\frac{\mu V_0}{12\hbar^2} & \int_{0}^{-b} ds \, \big[\delta^{(2)}(s+a)-\delta^{(2)}(s+b)\big] \nonumber \\
&\times \int_{0}^{\zeta} dw \, w^3 \, \tilde{T}^{W}_{B,I}(s,w) \nonumber \\
+\frac{\mu V_0}{12\hbar^2} & \int_{-b}^{-a} ds \,\big[\delta^{(2)}(s+a)-\delta^{(2)}(s+b)\big] \nonumber \\
&\times \int_{0}^{\zeta} dw \, w^3 \, \tilde{T}^{W}_{B,II}(s,w) \nonumber \\
+\frac{\mu V_0}{12\hbar^2} & \int_{-a}^{\eta} ds \,\big[\delta^{(2)}(s+a)-\delta^{(2)}(s+b)\big] \nonumber \\
&\times \int_{0}^{\zeta} dw \, w^3 \, \tilde{T}^{W}_{B,III}(s,w). \label{t10uvfinbbar13}
\end{align}
where, the kernel pieces $\tilde{T}^{W}_{B,I}$, $\tilde{T}^{W}_{B,II}$, and $\tilde{T}^{W}_{B,III}$ are defined by Eqs. (\ref{eq:barrierTKFweyl1}) - (\ref{eq:barrierTKFweyl3}), respectively. Substitution of these values and performing integration by parts give
\begin{equation}\label{t10uvfinbbar132}
\begin{split}
\tilde{T}_{B,III}^{S,(1)}(\eta,\zeta)&=0.
\end{split}
\end{equation}
Hence, it turns out that all pieces of the leading TKF correction vanishes, and consequently 
\begin{equation}
    \bar{\tau}^{S,(1)}_{B} =0 .
\end{equation}

We now generalize our analysis to arbitrary $n$th order terms $\bar{\tau}^{S,(n)}_{B}$, which follows from the calculation of $\tilde{T}_{B}^{S,(1)}(\eta,\zeta)$. We divide the integral along $s$ into different pieces corresponding to the regions where $\eta$ may fall. The $n$th order quantum corrections to the three pieces of the Weyl-quantized TKF are given below
\begin{widetext}
\begin{equation}\label{tnuvfinetan1}
\begin{split}
\tilde{T}_{B,I}^{S,(n)}(\eta,\zeta)=\left(\frac{\mu }{\hbar^2}\right)\sum_{r=1}^{n} \frac{1}{(2r+1)!}\frac{1}{2^{2r}}\int_{0}^{\eta} ds \,V^{(2r+1)}\left(s\right)\int_{0}^{\zeta} dw \, w^{2r+1} \, \tilde{T}_{B,I}^{S,(n-r)}(s,w),
\end{split}
\end{equation}
\begin{equation}\label{tnuvfinetan2}
\begin{split}
\tilde{T}_{B,II}^{S,(n)}(\eta,\zeta)=&\left(\frac{\mu }{\hbar^2}\right)\sum_{r=1}^{n} \frac{1}{(2r+1)!}\frac{1}{2^{2r}}\int_{0}^{-b} ds \,V^{(2r+1)}\left(s\right)\int_{0}^{\zeta} dw \, w^{2r+1} \, \tilde{T}_{B,I}^{S,(n-r)}(s,w)\\
&+\left(\frac{\mu }{\hbar^2}\right)\sum_{r=1}^{n} \frac{1}{(2r+1)!}\frac{1}{2^{2r}}\int_{-b}^{\eta} ds \,V^{(2r+1)}\left(s\right)\int_{0}^{\eta} dw \, w^{2r+1} \, \tilde{T}_{B,II}^{S,(n-r)}(s,w),
\end{split}
\end{equation}
\begin{equation}\label{tnuvfinetan3}
\begin{split}
\tilde{T}_{B,III}^{S,(n)}(\eta,\zeta)=&\left(\frac{\mu }{\hbar^2}\right)\sum_{r=1}^{n} \frac{1}{(2r+1)!}\frac{1}{2^{2r}}\int_{0}^{-b} ds \,V^{(2r+1)}\left(s\right)\int_{0}^{\zeta} dw \, w^{2r+1} \, \tilde{T}_{B,I}^{S,(n-r)}(s,w)\\
&+\left(\frac{\mu }{\hbar^2}\right)\sum_{r=1}^{n} \frac{1}{(2r+1)!}\frac{1}{2^{2r}}\int_{-b}^{-a} ds \,V^{(2r+1)}\left(s\right)\int_{0}^{\zeta} dw \, w^{2r+1} \, \tilde{T}_{B,II}^{S,(n-r)}(s,w)\\
&+\left(\frac{\mu }{\hbar^2}\right)\sum_{r=1}^{n} \frac{1}{(2r+1)!}\frac{1}{2^{2r}}\int_{-a}^{\eta} ds \,V^{(2r+1)}\left(s\right)\int_{0}^{\zeta} dw \, w^{2r+1} \, \tilde{T}_{B,III}^{S,(n-r)}(s,w).
\end{split}
\end{equation}
\end{widetext}
where $\tilde{T}_{B,l}^{S,(n-r)}$, for $l=I,II,III$ is the $l$th piece of the kernel $\tilde{T}_{B}^{S,(n-r)}$ in the $l$th region. The factor $\tilde{G}(s,w)$ in Eq. (\ref{gsw}) simply evaluates to unity so it no longer appears in the above equations. Also, the factor $V^{(2r+1)}\left(s\right)$ is the $(2r+1)$th derivative of the interaction potential defined in Eq. (\ref{potinheavd3}).

Eqs. (\ref{tnuvfinetan1}) - (\ref{tnuvfinetan3}) are recurrence relations whose initial conditions are the three pieces of the Weyl-quantized TKF defined in Eqs. (\ref{eq:barrierTKFweyl1}) - (\ref{eq:barrierTKFweyl3}). Evaluating the above integrals by performing integration by parts $2r-$times along $s$, using the initial conditions, and imposing the vanishing of the three pieces of the leading TKF correction, $\tilde{T}_{B,I}^{S,(1)}(\eta,\zeta) = \tilde{T}_{B,II}^{S,(1)}(\eta,\zeta) = \tilde{T}_{B,III}^{S,(1)}(\eta,\zeta) =0$, we finally obtain the relation
\begin{equation}\label{nthcorrectbarrier}
\tilde{T}_{B,I}^{S,(n)}(\eta,\zeta) = \tilde{T}_{B,II}^{S,(n)}(\eta,\zeta) = \tilde{T}_{B,III}^{S,(n)}(\eta,\zeta) =0. 
\end{equation}
Consequently, all the quantum corrections to the Weyl-quantized TKF vanish, that is, 
\begin{equation}
\bar{\tau}^{S,(n)}_{B} =0.
\end{equation}
This ultimately suggests that the Weyl-quantized TOA operator coincides with the supraquantized TOA operator for the case of a rectangular potential barrier, and that the tunneling time is still instantaneous
\begin{align}
\Delta\bar{\tau}^S=\Delta\bar{\tau}^W  . 
\end{align}

One might argue that this should have already been expected from the start since for each region, the potentials are linear, with either $V(q)=0$ or $V(q)=V_0$. However, we should be careful since there are, in fact, jump discontinuities at the edges of the barrier which may introduce cross terms among the different pieces of the Weyl-quantized barrier TOA operator. And since the full functional form of the supraquantized TOA operator was not available before, the Weyl-quantized barrier TOA operator is not immediately considered an algebra-preserving TOA operator. But as we have shown here, the quantum corrections do not contribute. Now, the vanishing of the TKFs $\tilde{T}^{S,(n)}_B$ imply that the expected barrier traversal time we have derived using Weyl-quantization still holds even if we impose TE-CCR.

\section{Tunneling time using the TOA eigenfunctions}
\label{sec:eigs}

We now tackle the problem anew using the momentum-space representation of the TOA-operators and TOA-eigenfunctions to demonstrate the full extent of the predicted instantaneous tunneling time. The TOA-operators can be expressed as the first moment of the identity generated by the TOA-eigenfunctions    
\begin{equation}
\mathsf{\hat{T}} = \int_{-\infty}^\infty d\tau  |\tau\rangle\langle\tau|\tau,  
\label{eq:toa-moment}
\end{equation}
where, $|\tau\rangle$ is an eigenket of the TOA-operator. It is also easy to see that Eq. \eqref{eq:toa-moment} satisfies time-reversal symmetry $\mathsf{\hat{\Theta}} \mathsf{\hat{T}} \mathsf{\hat{\Theta}^{-1}} = -\mathsf{\hat{T}}$, i.e., $\mathsf{\hat{\Theta}}|\tau\rangle=|-\tau\rangle$. It follows that the TOA difference Eq. \eqref{eq:toadiff0} now takes the form\footnote{In the succeeding expressions, we will drop the superscripts `W', `Q' and `S' since we have shown in Sec. \ref{sec:tunnelelingall} that $\mathsf{\hat{T}^W}=\mathsf{\hat{T}^Q}=\mathsf{\hat{T}^S}$.}
\begin{align}
    \Delta\bar{\tau} =& \int_{-\infty}^\infty dp \, \psi^*(p) \int_{-\infty}^\infty dp' \,  \psi(p')  \int_{-\infty}^\infty d\tau \, \tau   \nonumber \\
    & \times  \left\{ \Phi_F(\tau,p)\Phi_F^*(\tau,p') - \Phi_{B}(\tau,p)\Phi_{B}^*(\tau,p')\right\}, \label{eq:toadiff}
\end{align}
in which, $\Phi_F(\tau,p)$ and $\Phi_B(\tau,p)$ are the momentum-space representation of the free and barrier TOA-eigenfunctions, respectively. 

\subsection{Free TOA eigenfunctions}

Since the Weyl-ordered free TKF $T_F(\eta)$ coincides with the RHS extension of the Aharonov-Bohm TOA-operator, then it follows that $\mathsf{\hat{T}_{F}}=\mathsf{\hat{T}^{AB}}$ which admits the following eigenvalue equation
\begin{align}
\mathsf{\hat{T}_{F}}\Phi_F(\tau,p)=-i\mu\hbar \left( \dfrac{1}{p}\dfrac{d}{dp} -\dfrac{1}{p^2} \right) \Phi_F(\tau,p) = \tau \Phi_F(\tau,p). 
\end{align}
The eigenfunctions of $\mathsf{\hat{T}_{AB}}$ are already known \cite{giannitrapani1997positive}, which were obtained by treating $\Phi_F(\tau,p)$ as distributions and solved for the cases $p>0$ and $p<0$, i.e., $\mathsf{\hat{T}^{AB}}\Phi_F^{(\pm)}(\tau,p)=\tau\Phi_F^{(\pm)}(\tau,p)$. An alternative set of eigenfunctions called non-nodal and nodal eigenfunctions are then constructed by taking the sums and difference of $\Phi_F^{(\pm)}(\tau,p)$ and have the form 

\begin{subequations}
\begin{align}
    \Phi_{F}^{(non)}(\tau,p) =& \dfrac{1}{\sqrt{2\pi\hbar}}\sqrt{\dfrac{\abs{p}}{2\mu}} \exp[\dfrac{i}{\hbar} \dfrac{p^2}{2\mu}\tau], \label{eq:Feigsnon}\\
    \Phi_F^{(nod)}(\tau,p) =& \dfrac{1}{\sqrt{2\pi\hbar}}\sqrt{\dfrac{\abs{p}}{2\mu}} \exp[\dfrac{i}{\hbar} \dfrac{p^2}{2\mu}\tau] \text{sgn}(p) \label{eq:Feigsnod}.
\end{align}
\end{subequations}
The non-nodal TOA-eigenfunctions physically correspond to particle arrival with detection and contains the position-density of the time-evolved wavefunction, while the nodal TOA-eigenfunctions physically correspond to particle arrival without detection \cite{sombillo2016particle}. This now motivates us to also obtain the eigenfunctions $\Phi_B^{(non)}(\tau,p)$ and $\Phi_B^{(nod)}(\tau,p)$, and postulate that the same physical interpretation holds.

\begin{figure}[t!]
    \centering
    \begin{subfigure}
        \centering
        \includegraphics[width=0.45\textwidth]{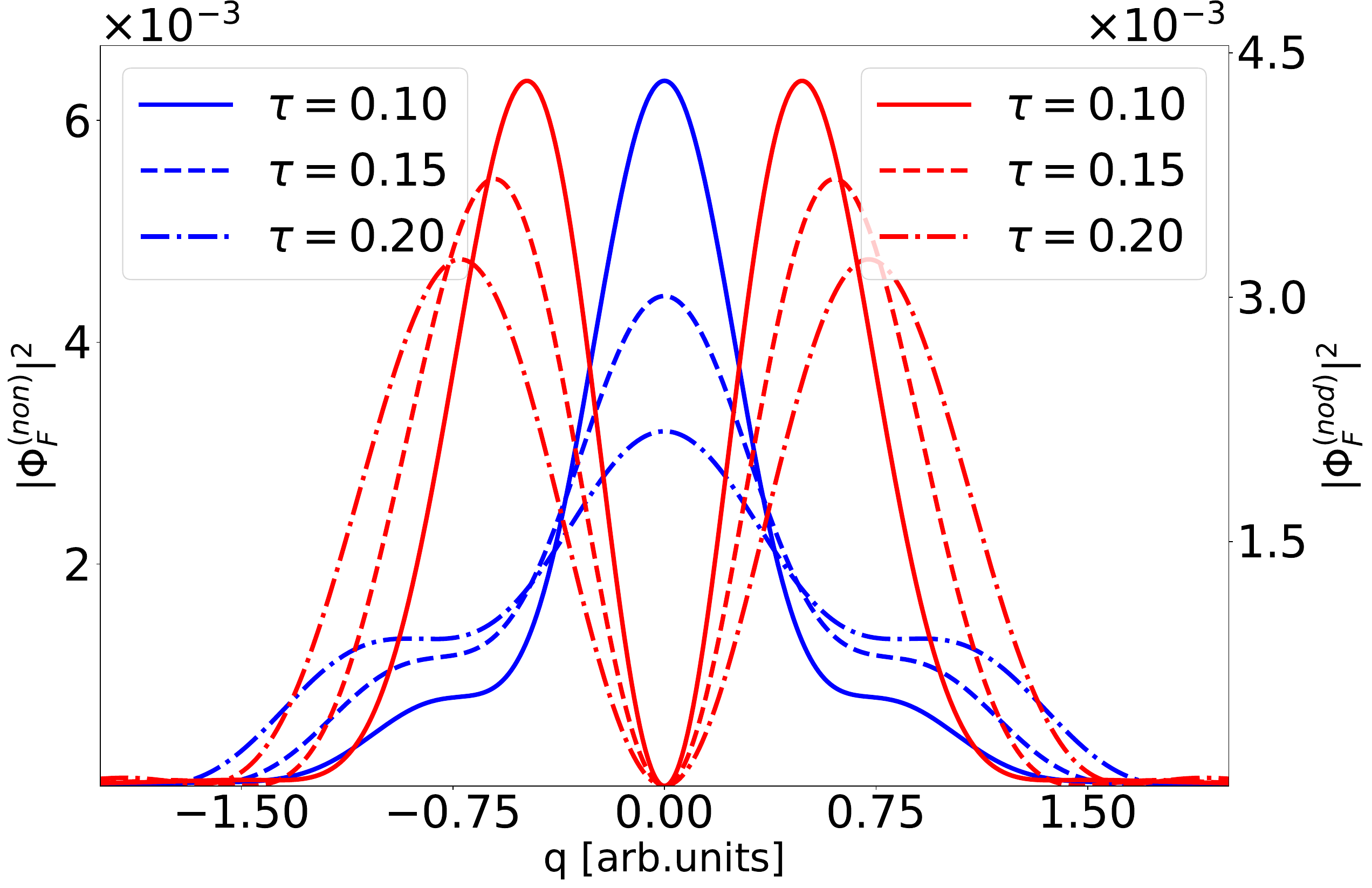}
    \end{subfigure} \hfil
    \begin{subfigure}
        \centering
        \includegraphics[width=0.45\textwidth]{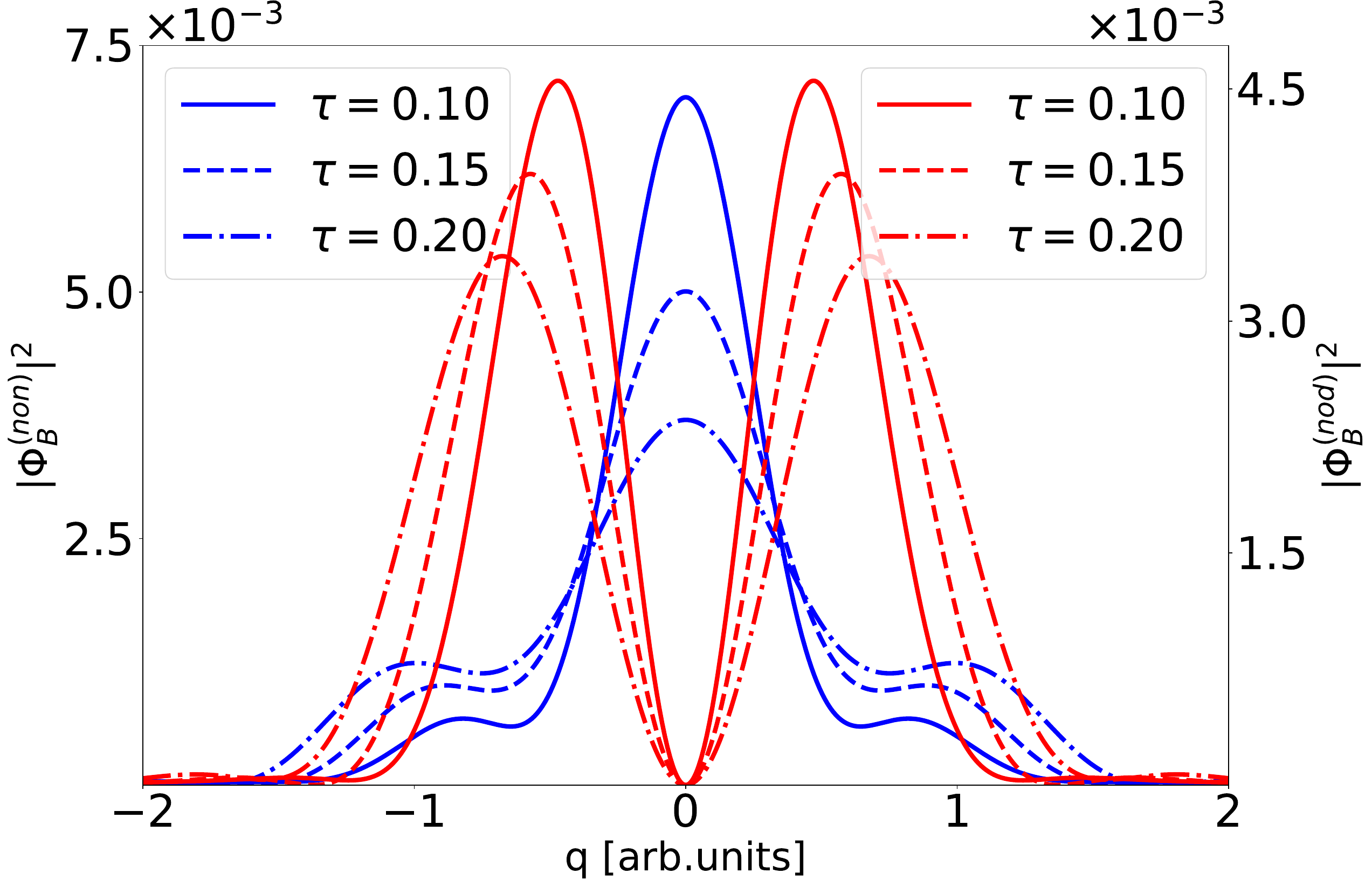}
    \end{subfigure} \hfil
    \caption{Position density distribution of the non-nodal and nodal eigenfunctions $\Phi_F(\tau,p)$ (top) and $\Phi_B(\tau,p)$ (bottom) for the parameters $\mu=\hbar=V_0=1$,$L=0.5$, and $\epsilon=0.05$.}
    \label{fig:eigs}
\end{figure}	

\subsection{Barrier TOA eigenfunctions}

The TOA-operator $\mathsf{\hat{T}_B}$ corresponding to the TKF $\tilde{T}_{B,III}(\eta,\zeta)$ has the form  
\begin{align}
\mathsf{\hat{T}_B} =& -\dfrac{\mu}{2} \bigg( \mathsf{\hat{p}^{-1}} ( \mathsf{\hat{q}} + L ) + ( \mathsf{\hat{q}} + L ) \mathsf{\hat{p}^{-1}}  \bigg) \nonumber \\
&+ \mu L \mathsf{\hat{p}^{-1}} ( 1-2\mu V_o \mathsf{\hat{p}^{-2}}  )^{-1/2} H( |\mathsf{\hat{p}}| - \sqrt{2\mu V_o}),
\label{eq:TB}
\end{align}
where, $H(z)$ is the heaviside function (see Appendix \ref{app:verify} for details). The first term of Eq. \eqref{eq:TB}, 
\begin{align}
\mathsf{\hat{T}_B^{(1)}} = -\dfrac{\mu}{2} \bigg( \mathsf{\hat{p}^{-1}} ( \mathsf{\hat{q}} + L ) + ( \mathsf{\hat{q}} + L ) \mathsf{\hat{p}^{-1}}  \bigg),
\end{align}
is a free TOA for the interaction-free region (see Fig. \ref{fig:setup}). Meanwhile, the second term of Eq. \eqref{eq:TB}, 
\begin{align}
\mathsf{\hat{T}_B^{(2)}} = \mu L \mathsf{\hat{p}^{-1}} ( 1-2\mu V_o \mathsf{\hat{p}^{-2}}  )^{-1/2} H( |\mathsf{\hat{p}}| - \sqrt{2\mu V_o}),
\end{align}
is the traversal across the barrier region. Furthermore, Eq. \eqref{eq:TB} now makes it clear why $\mathsf{\hat{T}_B}$ is the same for all operator ordering rules discussed in Sec. \ref{sec:tunnelingQ}, i.e., $\mathsf{\hat{T}_B^{(1)}}$ is analogous to $\mathsf{\hat{T}_F}$ and is the same for all operator ordering rules while $\mathsf{\hat{T}_B^{(2)}}$ is purely a function of the momentum operator. For below-barrier traversals, Eq. \eqref{eq:TB} shows that the $\mathsf{\hat{T}_B}$ is simply a free TOA-operator since the second term $\mathsf{\hat{T}_B^{(2)}}$ vanishes because of the heaviside function, which then results to the instantaneous tunneling time discussed in Sec. \ref{sec:tunnelelingall}. We emphasize that the vanishing of the below-barrier components was never imposed during the construction of the TKF $\tilde{T}_{B,III}(\eta,\zeta)$ in Sec. \ref{sec:toaopr}, and that instantaneous tunneling time naturally arose from the formulation.

The barrier TOA-eigenvalue problem in momentum representation is now
\begin{align}
\mathsf{\hat{T}_{B}} \Phi_B^{(\pm)}(\tau,p) =& -\dfrac{\mu}{2} \left( \dfrac{i\hbar}{p} \dfrac{d}{dp} + i\hbar \dfrac{d}{dp} \dfrac{1}{p} \dfrac{2L}{p} \right) \Phi_B^{(\pm)}(\tau,p) \nonumber \\
& +\dfrac{\mu L}{p} \left( 1 - \dfrac{2 \mu V_o}{p^2}\right)^{-1/2} \Phi_B^{(\pm)}(\tau,p)
\end{align}
for $|p| > 2\mu V_o$, and 
\begin{align}
\mathsf{\hat{T}_{B}} \Phi_B^{(\pm)}(\tau,p) =& -\dfrac{\mu}{2} \left( \dfrac{i\hbar}{p} \dfrac{d}{dp} + i\hbar \dfrac{d}{dp} \dfrac{1}{p} + \dfrac{2L}{p} \right) \Phi_B^{(\pm)}(\tau,p)
\end{align}
for for $|p| < 2\mu V_o$. The eigenfunctions $\Phi_B^{(\pm)}(\tau,p)$ are again treated as distributions and solved separately for the cases $p>0$ and $p<0$, which are then used to construct the non-nodal and nodal eigenfunctions which have the form 
\begin{subequations}
\begin{align}
\Phi_B^{(non)}(\tau,p) =& \dfrac{1}{\sqrt{2\pi\hbar}}\sqrt{\dfrac{\abs{p}}{2\mu}} \exp[\dfrac{i}{\hbar} \dfrac{p^2}{2\mu}\tau] \exp[\dfrac{i}{\hbar} |p| L] f(p) \label{eq:Beigsnon}
\end{align}
\begin{align}
\Phi_B^{(nod)}(\tau,p) =& \Phi_B^{(non)}(\tau,p) \text{sgn}(p) \label{eq:Beigsnod}
\end{align}
\begin{align}
f(p)=& 
	\begin{cases}
		\exp[-\dfrac{i \abs{p}L}{\hbar} \sqrt{1 - \dfrac{2 \mu V_o}{p^2}}] \quad ,& \abs{p}\geq\sqrt{2 \mu V_o}\\
		1 \quad ,& \abs{p}<\sqrt{2 \mu V_o}
	\end{cases} \label{eq:factor}
\end{align}
\end{subequations}
It is easy to see that $\Phi_B^{(non)}(\tau,p)$ and $\Phi_B^{(nod)}(\tau,p)$ reduces to $\Phi_F^{(non)}(\tau,p)$ and $\Phi_F^{(nod)}(\tau,p)$ in the limit $V,L\rightarrow0$. Moreover, $\Phi_B(\tau,p)$ form a complete set of eigenfunctions, i.e., 
\begin{align}
\sum_{\alpha=\text{non},\text{nod}} & \int_{-\infty}^\infty d\tau \Phi_B^{(\alpha)*}(\tau,p)\Phi_B^{(\alpha)}(\tau,p') \nonumber \\
=& \dfrac{1}{2}\sqrt{\dfrac{|p|}{|p'|}} \exp[\dfrac{i}{\hbar}\big(|p|-|p'|\big)L]f^*(p)f(p') \nonumber \\
&\times \big\{ 1+(\text{sgn}p)(\text{sgn}p') \big\} \big\{ \delta(p-p') + \delta(p+p') \big\} \nonumber \\
=& \delta(p-p')
\end{align}
The last line follows from the fact that the factor $\{ 1+(\text{sgn}p)(\text{sgn}p') \}$ vanishes when $p$ and $p'$ have opposite signs, therefore, the contribution of $\delta(p+p')$ vanishes.

The position density distribution of the non-nodal and nodal eigenfunctions are centered at the origin $q=0$ which is the arrival point. The former has the common property of having a single-peak at the arrival point $q=0$ while the latter has two peaks that are symmetric around the origin and identically vanishes at the arrival point \cite{galapon2018quantizations}. This is shown in Fig. \ref{fig:eigs} by taking the Fourier transform of $\Phi(\tau,p)$, i.e., 
\begin{align}
\tilde{\Phi}(\tau,q) = \int_{-\infty}^\infty \dfrac{dp}{\sqrt{2\pi\hbar}} e^{ipq/\hbar} \Phi(\tau,p). 
\end{align}
In anticipation that the integral may diverge, we insert a converging factor $\lim_{\epsilon\rightarrow0}e^{-\epsilon p^2}$.

\subsection{Time-of-arrival difference}

We now use the TOA-eigenfunctions $\Phi_{F/B}(\tau,p)$ to calculate the tunneling time. Direct substitution of Eqs. \eqref{eq:Feigsnon}-\eqref{eq:Feigsnod} and Eqs. \eqref{eq:Beigsnon}-\eqref{eq:factor} into Eq. \eqref{eq:toadiff} yields
\begin{align}
\Delta\bar{\tau} 
=& \dfrac{\mu\hbar}{2i} \int_{-\infty}^\infty\int_{-\infty}^\infty dp dp' \psi^*(p)\psi(p') \dfrac{\sqrt{|p|}}{p}\dfrac{\sqrt{|p'|}}{p'} \nonumber \\
&\times \left\{ 1 - \exp[\dfrac{i}{\hbar}\left(|p|-|p'|\right)L] f(p)f^*(p') \right\}\nonumber \\
&\times  \left\{1+(\text{sgn}p)(\text{sgn}p')\right\} \dfrac{d}{dp} \big\{ \delta(p-p') + \delta(p+p') \big\}.
\end{align}
Notice that the factor $\{1+(\text{sgn}p)(\text{sgn}p')\}$ constrains the integration region to $\int_0^\infty\int_0^\infty dpdp' + \int_{-\infty}^0\int_{-\infty}^0dpdp'$, therefore, the contribution of the term $\delta(p+p')$ vanishes. Let us now evaluate the term corresponding to the region $\int_0^\infty\int_0^\infty dpdp'$, i.e., 
\begin{align}
\Delta\bar{\tau}_{(+)} = \dfrac{\mu\hbar}{i} \int_{0}^\infty dp &  \dfrac{\psi^*(p)}{\sqrt{p}} \dfrac{d}{dp}\int_{0}^\infty dp'  \dfrac{\psi(p')}{\sqrt{p'}} \delta(p-p') \nonumber \\
-  \dfrac{\mu\hbar}{i}  \int_{0}^\infty & dp  \dfrac{\psi^*(p)}{\sqrt{p}} \exp[\dfrac{i}{\hbar}pL]f(p) \nonumber \\
\dfrac{d}{dp}  \int_{0}^\infty & dp'  \dfrac{\psi(p')}{\sqrt{p'}} \exp[-\dfrac{i}{\hbar}p'L]f^*(p') \delta(p-p') \nonumber \\ 
= \mu L \int_0^\infty dp & \dfrac{|\psi(p)|^2}{p} \nonumber \\
-\dfrac{\mu\hbar}{i}  \int_0^\infty & dp \dfrac{|\psi(p)|^2}{p} f(p)\dfrac{df^*(p)}{dp}. 
\end{align}
Repeating the same steps for the region $\int_{-\infty}^0\int_{-\infty}^0dpdp'$, and adding the results, we finally obtain 
\begin{align}
\Delta\bar{\tau} =& \text{P.V.}\left[ \int_{-\infty}^\infty dp \dfrac{\mu L }{p} |\psi(p)|^2 \right] \nonumber \\
&- \int_{\sqrt{2\mu V_o}}^\infty dp\dfrac{\mu L}{\sqrt{p^2-2\mu V_o}} \big( |\psi(p)|^2 - |\psi(-p)|^2 \big),     
\end{align}
which is exactly equal to the $\Delta\bar{\tau}^W$ defined in Eqs. \eqref{eq:toadiff2}-\eqref{eq:R}. This now leads us to hypothesize that instantaneous tunneling time is a physical manifestation of the completeness of the TOA-eigenfunctions, as well as the time-reversal symmetry satisfied by the TOA-operator.

\section{Free vs Barrier TOA distributions}
\label{sec:toadist}

The TOA difference Eq. \eqref{eq:toadiff0} only provides the average TOA measurement after repeatedly performing the experiment for an ensemble of identically prepared particles. If tunneling time is indeed instantaneous, then the corresponding TOA-distribution of an incident wavepacket with only below-barrier components should indicate a shift of the TOA-distribution at earlier times compared to a free particle as if the barrier was not even present. The probability that a particle prepared in the state $\ket{\psi}$ will arrive at $q=0$ at a time $t$ before $\tau$ is 
\begin{equation}
\langle \psi | \mathsf{\hat{\Pi}} | \psi \rangle = \int_{-\infty}^\tau dt \braket{\psi}{t}\braket{t}{\psi},     
\end{equation}
wherein $\mathsf{\hat{\Pi}}$ is the positive operator valued measure associated to the TOA-operator, and $\ket{\tau}$ is the TOA eigenvector with eigenvalue $\tau$. The TOA-distribution is then constructed from
\begin{align}
\Pi_\psi(\tau) = \frac{d}{d\tau} \langle \psi | \mathsf{\hat{\Pi}} | \psi \rangle = \left| \langle \psi | \tau \rangle \right|^2
\end{align}
which is equivalent to the overlap of the incident wavepacket with the TOA-eigenfunction.

\begin{figure}[t!]
	\centering
	\includegraphics[width=0.45\textwidth]{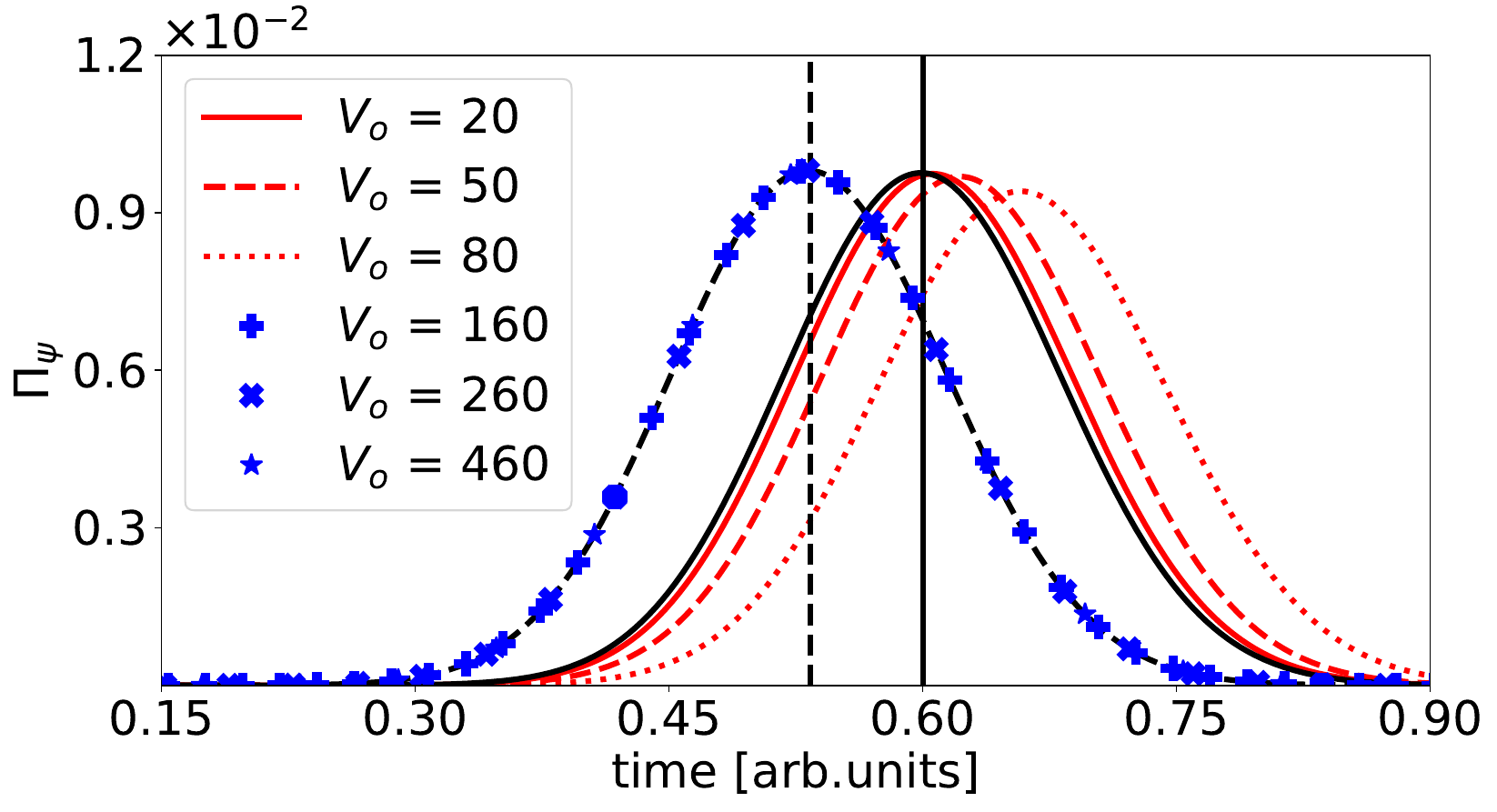}
	\caption{Comparison of the TOA distributions in the absence and presence of a barrier for a Gaussian wavepacket with only above barrier (red) and below barrier (blue) components initially centered at $q_o=-9$ with momentum $k_o=15$ and position variance $\sigma=1.2$ for the parameters $\mu=\hbar=1$. The solid black TOA distribution indicates the absence of a potential barrier while the dashed black TOA distribution indicates only the interaction-free region in the presence of a barrier (see Fig. \ref{fig:setup}). }
	\label{fig:DistCompare}
\end{figure}

To illustrate, let us now consider a single Gaussian wavepacket $\psi(q)=(\sigma \sqrt{2\pi})^{-1/2}e^{-(q-q_o)^2/4\sigma^2}e^{i k_o q}$ and construct the corresponding TOA distribution for various barrier heights as shown in Fig. (\ref{fig:DistCompare}). It can be seen that if the barrier height $V_o$ is low such that the support of $\tilde{\psi}(k)$ is above $\kappa_o$, then the presence of the barrier will on average cause a delay in the TOA compared to the free case. This is consistent with the known classical dynamics of a particle, i.e., the particle loses kinetic energy as it traverses above the barrier resulting to lower velocities and larger TOA. Meanwhile, if the barrier height $V_o$ is high such that the support of $\tilde{\psi}(k)$ is below $\kappa_o$, then the barrier will on average cause an earlier arrival at $q=0$ compared to the free case, due to instantaneous tunneling time. The manifestation of instantaneous tunneling time thus corresponds to a shift in the peak of the TOA distribution of a tunneled particle to earlier times compared to a free particle, and is equivalent to a free particle that only traveled the interaction-free regions, highlighted in blue in Fig. \ref{fig:setup}, as if the barrier was not even present. However, it can also be seen that the TOA distribution spreads to values that indicate finite tunneling times which arise from the dispersion of the wavepacket as it propagates to arrive at $q=0$.  

Our results align with those reported in Refs. \cite{dias2017space,de2024traveling}, which employ a self-adjoint time operator canonically conjugate to the Hamiltonian through an alternative quantization rule. Specifically, we observe similar behavior to Fig. 2 of Ref. \cite{dias2017space} and note consistency with the instantaneous tunneling described in Ref. \cite{de2024traveling}, where only energy above the barrier influences the mean arrival time.

\section{Conclusions}
\label{sec:conc}

We have investigated the tunneling time problem using the theory of TOA-operators by calculating the tunneling time for all the possible quantum images, $\mathsf{\hat{T}^{Q}}$ and $\mathsf{\hat{T}^{S}}$, of the classical TOA that were not considered in Ref. \cite{galapon2012only} to gain insights on the role of operator ordering rules and the TE-CCR to the predicted instantaneous tunneling time. The TOA-operators $\mathsf{\hat{T}^{Q}}$ are constructed by canonical quantization of the classical TOA and generally do not satisfy the TE-CCER. Meanwhile, the TOA-operators $\mathsf{\hat{T}^{S}}$ is constructed via \textit{supraquantization} to ensure that it satisfies the TE-CCR. We have shown that instantaneous tunneling time persists for all these TOA-operators, i.e., neither operator ordering rules or the TE-CCR play a role in the predicted instantaneous tunneling time. The full extent of the instantaneous tunneling time is then demonstrated by comparing the TOA-distributions of a Gaussian wavepacket in the absence and presence of a square barrier. We have shown that the peak of the wavepacket that tunneled is shifted to earlier times than a free wavepacket, and that it is exactly equal $L/\nu_o$, where $L$ is the barrier length and $\nu_o$ is the group velocity. 

Although it is still not clear whether there is a preferred mapping between classical observables and its quantum image, our results suggest that if we assume different ordering rules to correspond to different experimental situations \cite{ali1990ordering}, then quantum tunneling is strictly an instantaneous quantum phenomenon, regardless of the characteristics of experimental setup being considered. We have also recently introduced the concept of partial-tunneling and full-tunneling processes across an arbitrary potential barrier $V(q)$ in Ref. \cite{flores2024partial} to explain the seemingly contradictory non-zero and vanishing tunneling times using only the Weyl-ordered TOA-operators. 

To illustrate, consider two square barriers with heights $V_1$ and $V_2$ where $V_1<V2$ and use the same measurement scheme shown in Fig. \ref{fig:setup}. Full-tunneling happens when all the energy components are below $V_1$ and tunneling time is instantaneous. Partial tunneling happens when the energy components are above $V_1$ but below $V_2$, which means that the wavepacket traversed above $V_1$ but tunneled through $V_2$. We have argued that the time associated to this should not be interpreted as a tunneling time because this 'time' comes from the traversal above $V_1$. One can then generalize this by by approximating any barrier as a series of square barriers with varying heights and widths and take the continuous limit. 

Applying the operators $\mathsf{\hat{T}^{Q}}$ and $\mathsf{\hat{T}^{S}}$ to extract the time associated to a partial tunneling process may then provide new insights on the problem of operator ordering. Specifically, we have seen in Eq. \eqref{eq:TB} that the second term of $\mathsf{\hat{T}_B}$ is purely a function of the momentum operator for a square potential barrier. For arbitrary potential barriers $V(q)$, the second term of Eq. \eqref{eq:TB} will be different for each $\mathsf{\hat{T}^{Q}}$ because of the different ordering rules between the position and momentum operators. Similarly, $\mathsf{\hat{T}^{S}}$ will have non-zero correction terms that will arise due to the TE-CCR. Thus, the time associated to partial-tunneling will be different for all operators $\mathsf{\hat{T}^{Q}}$ and $\mathsf{\hat{T}^{S}}$. We leave this problem open for future studies.


\appendix

\begin{widetext}

\section{Canonical quantization of the classical TOA for arbitary ordering rules}
\label{app:deformation}

For completeness, we provide the full details on the derivation of the TOA-operators Eqs. \eqref{eq:toaRHS}-\eqref{eq:deformation}. In coordinate representation, the TOA-operators have the form 
\begin{align}
    (\mathsf{\hat{T}^Q}\phi)(q) =&  \int_{-\infty}^\infty dq'  \langle q | \mathsf{\hat{T}}  | q' \rangle \phi(q') \nonumber \\
    =& \int_{-\infty}^\infty dq'  \sum_{m=0}^\infty  \dfrac{(2m-1)!! (-\mu)^{m+1}}{m! }  \sum_{n=1}^\infty a_n^{(m)} \langle q | \mathsf{\hat{t}^Q_{n,-2m-1}} | q' \rangle \phi(q').
    \label{eq:app-rhstoa}
\end{align}
where, $\mathsf{\hat{t}^Q_{n,-2m-1}}$ denotes a specific ordering on the quantization of the monomial $q^np^{-(2m+1)}$. The second line of Eq. \eqref{eq:app-rhstoa} follows directly from the TOA-operator expansion Eq. \eqref{eq:toaexpansion}. Substituting Eq. \eqref{eq:BDbasis} into Eq. \eqref{eq:app-rhstoa} yields  
\begin{align}
    (\mathsf{\hat{T}^Q}\phi)(q) =& \int_{-\infty}^\infty dq'  \sum_{m=0}^\infty  \dfrac{(2m-1)!! (-\mu)^{m+1}}{m! }  \sum_{n=1}^\infty a_n^{(m)} \sum_{j=0}^\infty \dfrac{\hbar^j \alpha_j^Q  \Gamma(-2m)\Gamma(n+1)}{\Gamma(-2m-j)\Gamma(n-j+1)} \langle q |  \mathsf{\hat{t}^{W}_{n-j,-2m-1-j}} | q' \rangle \phi(q') 
    \label{eq:app-rhstoa1}
\end{align}
in which, $\mathsf{\hat{t}^{W}_{n-j,-2m-1-j}}$ are the Weyl-ordered Bender-Dunne basis operators in Eqs. \eqref{eq:BDbasis}-\eqref{eq:coeff} while $\alpha_j^Q$ are the Taylor expansion coefficients of the ordering function $\Theta^{Q}(x)=\sum_j \alpha_j^Q x^j$. 

Let us now consider the kernel of the $\mathsf{\hat{t}^{W}_{n-j,-2m-1-j}}$, i.e.,
\begin{align}
\langle q |  \mathsf{\hat{t}^{W}_{n-j,-2m-1-j}} | q' \rangle =& \dfrac{1}{2^{n-j+1}} \sum_{k=0}^{n-j} \dfrac{(n-j)!}{(n-j-k)!k!} \langle q | \mathsf{\hat{q}}^k \mathsf{\hat{p}}^{-2m-1-j} \mathsf{\hat{q}}^{n-j-k} | q' \rangle \nonumber \\
=& \dfrac{1}{2^{n-j+1}} \sum_{k=0}^{n-j} \dfrac{(n-j)!}{(n-j-k)!k!} q^k q'^{n-j-k} \langle q | \mathsf{\hat{p}}^{-2m-1-j} | q' \rangle \nonumber \\
=& \dfrac{(q+q')^{n-j}}{2^{n-j+1}} \dfrac{i^{2m+1+j}}{2\hbar^{2m+1+j}(2m+j)!}(q-q')^{2m+j} \text{sgn}(q-q').
\label{eq:app-rhstoa2}
\end{align}
The last line directly follows from the identity Eq. \eqref{eq:momentumkernel}. Substituting Eq. \eqref{eq:app-rhstoa2} into Eq. \eqref{eq:app-rhstoa1}, and rearranging the summation, we get 
\begin{align}
(\mathsf{\hat{T}^Q}&\phi)(q) \nonumber \\
%
%
%
%
=&  \dfrac{1}{2} \int_{-\infty}^\infty dq' \phi(q') \text{sgn}(q-q') \sum_{m=0}^\infty \dfrac{(2m-1)!!}{(2m)!m!}   \left( \dfrac{\mu}{\hbar^2} (q-q')^2 \right)^m  \sum_{n=1}^\infty a_n^{(m)}  \sum_{j=0}^\infty \alpha_j^Q  (-i)^j (q-q')^j \dfrac{\Gamma(n+1)}{\Gamma(n-j+1)} \left(\dfrac{q+q'}{2}\right)^{n-j} \nonumber \\
=&  \dfrac{1}{2} \int_{-\infty}^\infty dq' \phi(q') \text{sgn}(q-q') \sum_{j=0}^\infty \alpha_j^Q  \left(-i\zeta\dfrac{d}{d\eta}\right)^j  \sum_{m=0}^\infty \dfrac{(2m-1)!!}{(2m)!m!}   \left( \dfrac{\mu}{\hbar^2} \zeta^2 \right)^m  \sum_{n=1}^\infty a_n^{(m)}  \eta^n \nonumber \\
=&  \dfrac{1}{2} \int_{-\infty}^\infty dq' \phi(q') \text{sgn}(q-q') \sum_{j=0}^\infty \alpha_j^Q  \left(-i\zeta\dfrac{d}{d\eta}\right)^j  \sum_{m=0}^\infty \dfrac{(2m-1)!!}{(2m)!m!}   \left( \dfrac{\mu}{\hbar^2} \zeta^2 \right)^m \int_0^\eta ds ( V(\eta) - V(s) )^m, 
\label{eq:app-rhstoa3}
\end{align}
wherein, we performed the change of variables $\eta=(q+q')/2$ and $\zeta=q-q'$. The last line follows from the assumption that the potential $V(\eta)$ is analytic at the origin, see Eq. \eqref{eq:ltoa2}. Interchanging the order of summation and integration, then performing the necessary operations we finally obtain
\begin{subequations}
\begin{align}
    (\mathsf{\hat{T}^Q}\phi)(q) =& \int_{-\infty}^\infty dq' \dfrac{\mu}{i\hbar} \bigg[ \mathsf{\hat{\Theta}^Q}  \tilde{T}^W(\eta,\zeta) \bigg] \text{sgn}(q-q') \phi(q') \label{eq:apptoaRHS},\\
    \tilde{T}^W(\eta,\zeta) =&  \dfrac{1}{2} \int_0^\eta ds \, {_0}F_1\left[ ; 1; \dfrac{\mu}{2 \hbar^2}(V(\eta)-V(s))\zeta^2\right], \label{eq:app-weyltkf}\\
    \mathsf{\hat{\Theta}^Q} =& \Theta^Q \left( -i\zeta\frac{d}{d\eta} \right)= \sum_{j=0}^\infty \alpha_j^Q (-i\zeta)^j \dfrac{d^j}{d\eta^j} \label{eq:app-deformation} 
\end{align}    
\end{subequations}
in which, $T(q,q')=\tilde{T}(\eta,\zeta)$, and ${_0}F_1(;1;z)$ is a specific hypergeometric function.


\section{Examples of the TKF transformation form Weyl to non-Weyl}
\label{app:examples}

Ref. \cite{galapon2018quantizations} was only able to perform the quantization using Weyl, Born-Jordan, and simple symmetric ordering rule, which are given by\footnote{The superscripts, `BJ', and `SS' will be used to denote quantities related to Born-Jordan, and simple-symmetric ordering, respectively}
\begin{align}
T^{W}(q,q') =& \dfrac{1}{2} \int_0^{\frac{q+q'}{2}} ds \, {_0}F_1\left[ ; 1; \dfrac{\mu}{2 \hbar^2} (q-q')^2 \left\{ V\left(\dfrac{q+q'}{2}\right)-V(s) \right\}\right], \label{eq:nonrelWeyl}
\end{align}
\begin{align}
T^{BJ}(q,q') =& \dfrac{1}{2(q-q')} \int_{q'}^q ds \int_0^s du \,{_0}F_1\left[;1;\dfrac{\mu}{2\hbar^2}(q-q')^2 \left\{ V(s) - V(u)\right\}\right],\label{eq:nonrelBJ}
\end{align}
\begin{align}
T^{SS}(q,q')=& \dfrac{1}{4} \int_0^q ds \,  {_0}F_1\left[;1;\dfrac{\mu}{2\hbar^2}(q-q')^2 \left\{ V(q) - V(s)\right\}\right] + \dfrac{1}{4} \int_0^{q'} ds \, {_0}F_1\left[;1;\dfrac{\mu}{2\hbar^2}(q-q')^2 \left\{ V(q') - V(s)\right\}\right], \label{eq:nonrelSS} 
\end{align}
where ${_0}F_1(;1;z)$ is a particular hypergeometric function.

We shall use the linear potential $V(q)=\lambda q$ and harmonic oscillator potential $V(q)=\mu \omega^2 q^2 / 2$ to demonstrate the validity of the TKF transformation from Weyl to any ordering rule in Eqs. \eqref{eq:toaRHS}-\eqref{eq:weyltkf}. The ordering function associated to the Born-Jordan and simple-symmetric quantizations are $\text{sinc}(x/2)$ and $\cos(x/2)$, respectively. Thus, the deformation operator associated to these ordering rules are
\begin{subequations}
\begin{align}
\mathsf{\hat{\Theta}_{BJ}} =& \text{sinc}\left[-\dfrac{i\zeta}{2} \dfrac{d}{d\eta}\right]= \sum_{n=0}^\infty \dfrac{(-1)^n}{(2n+1)!} \left(\dfrac{-i\zeta}{2}\right)^{2n} \dfrac{d^{2n}}{d\eta^{2n}} \label{eq:deform_bj}\\
\mathsf{\hat{\Theta}_{SS}} =& \cos\left[-\dfrac{i\zeta}{2} \dfrac{d}{d\eta}\right] = \sum_{n=0}^\infty \dfrac{(-1)^n}{(2n)!} \left(\dfrac{-i\zeta}{2}\right)^{2n} \dfrac{d^{2n}}{d\eta^{2n}}\label{eq:deform_ss}
\end{align}  
\end{subequations}

\subsection{Linear potential}

Direct substitution of the potential $V(q)=\lambda q$ into Eqs. \eqref{eq:nonrelWeyl}-\eqref{eq:nonrelSS} yield the following TKFs
\begin{align}
\tilde{T}^W(\eta,\zeta) =& \dfrac{\eta}{2} {_0}F_1\left[;2;\dfrac{\mu \lambda}{2\hbar^2}\eta\zeta^2\right] \label{eq:Weyl_lin}    
\end{align}
\begin{align}
\tilde{T}^{BJ}(\eta,\zeta) =& \dfrac{1}{4\zeta} \left( \eta + \dfrac{\zeta}{2} \right)^2 {_0}F_1\left[;3;  \left( \dfrac{\mu \lambda}{2\hbar^2} \zeta^2 \right)  \left( \eta + \dfrac{\zeta}{2} \right) \right] -  \dfrac{1}{4\zeta} \left( \eta - \dfrac{\zeta}{2} \right)^2  {_0}F_1\left[;3;   \left( \dfrac{\mu \lambda}{2\hbar^2} \zeta^2 \right)  \left( \eta - \dfrac{\zeta}{2} \right) \right] \label{eq:BJ_lin}
\end{align}
\begin{align}
T^{SS}(q,q') =& \dfrac{1}{4}  \left( \eta + \dfrac{\zeta}{2} \right) {_0}F_1\left[;2; \left( \dfrac{\mu \lambda}{2\hbar^2} \zeta^2 \right)  \left( \eta - \dfrac{\zeta}{2} \right) \right] + \dfrac{1}{4}  \left( \eta - \dfrac{\zeta}{2} \right) {_0}F_1\left[;2;\dfrac{\mu \lambda}{2\hbar^2}(q-q')^2q'\right] \label{eq:SS_lin}
\end{align}
where ${_0}F_1[;n;z]$ is a particular hypergeometric function, and we performed a change of variable from $(q,q')$ to $(\eta,\zeta)-$coordinates. Taking the power series expansion of Eq. \eqref{eq:Weyl_lin} we obtain
\begin{align}
\tilde{T}^W(\eta,\zeta)=& \dfrac{1}{2}\sum_{m=0}^{\infty}\frac{\eta^{m+1}}{m!(2)_{m}}\left(\frac{\mu\lambda}{2\hbar^{2}}\zeta^{2}\right)^{m},
\label{eq:Weyl_lin1}
\end{align}
and applying the `deformation operator' $\mathsf{\hat{\Theta}_{BJ}}$ yields
\begin{align}
    \tilde{T}^{(BJ)}(\eta,\zeta)=&\frac{\eta}{2}\sum_{m=0}^{\infty}\frac{\Gamma(m+2)}{m!(2)_{m}}\left(\frac{\mu\lambda}{2\hbar^{2}}\zeta^{2}\eta\right)^{m}\sum_{n=0}^{\infty}\frac{(-1)^{n}}{(2n+1)!}\frac{1}{\Gamma(m-2n+2)}\left(\frac{-i\zeta}{2\eta}\right)^{2n} \nonumber \\
    =&\frac{1}{4\zeta}\left(\eta+\frac{\zeta}{2}\right)^{2}{}_{0}F_{1}\left[;3;\left(\frac{\mu\lambda}{2\hbar^{2}}\zeta^{2}\right)\left(\eta+\frac{\zeta}{2}\right)\right] - \frac{1}{4\zeta}\left(\eta-\frac{\zeta}{2}\right)^{2}{}_{0}F_{1}\left[;3;\left(\frac{\mu\lambda}{2\hbar^{2}}\zeta^{2}\right)\left(\eta-\frac{\zeta}{2}\right)\right],
\end{align}
which is equal to Eq. \eqref{eq:BJ_lin}. Similarly, applying the deformation operator $\mathsf{\hat{\Theta}_{SS}}$ into Eq. \eqref{eq:Weyl_lin1} we get
\begin{align}
    \tilde{T}^{(SS)}(\eta,\zeta)=&\frac{\eta}{2}\sum_{m=0}^{\infty}\frac{\Gamma(m+2)}{m!(2)_{m}}\left(\frac{\mu\lambda}{2\hbar^{2}}\zeta^{2}\eta\right)^{m}\sum_{n=0}^{\infty}\frac{(-1)^{n}}{(2n)!}\frac{1}{\Gamma(m-2n+2)}\left(\frac{-i\zeta}{2\eta}\right)^{2n} \nonumber \\
    =&\frac{1}{4}\left(\eta+\frac{\zeta}{2}\right){}_{0}F_{1}\left[;2;\left(\frac{\mu\lambda}{2\hbar^{2}}\zeta^{2}\right)\left(\eta+\frac{\zeta}{2}\right)\right] + \frac{1}{4}\left(\eta-\frac{\zeta}{2}\right){}_{0}F_{1}\left[;2;\left(\frac{\mu\lambda}{2\hbar^{2}}\zeta^{2}\right)\left(\eta-\frac{\zeta}{2}\right)\right], 
\end{align}
which is equal to Eq. \eqref{eq:SS_lin}.

\subsection{Harmonic oscillator}

Direct substitution of the potential $V(q)=\mu \omega^2 q^2 / 2$ into Eqs. \eqref{eq:nonrelWeyl}-\eqref{eq:nonrelBJ} yields the following TKFs
\begin{align}
\tilde{T}^W(\eta,\zeta)=&\dfrac{1}{2} \left( \dfrac{\mu \omega}{\hbar} \zeta \right)^{-1} \sinh\left[\frac{\mu\omega}{\hbar}\eta\zeta\right] \label{eq:tkfharmonicweyl}   
\end{align}
\begin{align}
\tilde{T}^{BJ}(\eta,\zeta)=& \left(\dfrac{\hbar}{\mu\omega}\right)^2 \dfrac{1}{2\zeta^3} \left\{ \cosh\left[ \dfrac{\mu\omega}{\hbar}\zeta \left(\eta+\dfrac{\zeta}{2}\right) \right] - \cosh\left[ \dfrac{\mu\omega}{\hbar}\zeta \left(\eta-\dfrac{\zeta}{2}\right) \right] \right\} \nonumber \\
=& \left( \dfrac{\mu\omega}{2\hbar}\zeta^2 \right)^{-1}  \sinh \left[ \dfrac{\mu \omega}{2\hbar}\zeta^2 \right] \tilde{T}^{(W)}(\eta,\zeta) \label{eq:tkfharmonicbj}
\end{align}
\begin{align}
\tilde{T}^{SS}(\eta,\zeta)=& \dfrac{\hbar}{\mu\omega}\dfrac{1}{4\zeta} \left\{ \sinh\left[ \dfrac{\mu\omega}{\hbar}\zeta \left(\eta+\dfrac{\zeta}{2}\right) \right] + \sinh\left[ \dfrac{\mu\omega}{\hbar}\zeta \left(\eta-\dfrac{\zeta}{2}\right) \right] \right\} \nonumber \\
 =& \cosh \left[ \dfrac{\mu \omega}{2\hbar}\zeta^2 \right] \tilde{T}^{(W)}(\eta,\zeta) \label{eq:tkfharmonicsymmetric}
\end{align}
wherein, we performed a change of variable from $(q,q')$ to $(\eta,\zeta)$-coordinates and used the following identities
\begin{align}
\cosh x - \cosh y =& 2 \sinh\left[\dfrac{x+y}{2}\right]\sinh\left[\dfrac{x-y}{2}\right] \nonumber \\
\sinh x + \sinh y =& 2 \sinh\left[\dfrac{x+y}{2}\right]\cosh\left[\dfrac{x-y}{2}\right] \nonumber 
\end{align}
to simplify the TKFs $\tilde{T}^{BJ}(\eta,\zeta)$ and $\tilde{T}^{SS}(\eta,\zeta)$. 

We now apply the deformation operator $\mathsf{\hat{\Theta}_{BJ}}$ into Eq. \eqref{eq:tkfharmonicweyl} which yields
\begin{align}
\tilde{T}^{BJ}(\eta,\zeta) =& \dfrac{1}{2} \left( \dfrac{\mu \omega}{\hbar} \zeta \right)^{-1}  \sum_{n=0}^\infty \dfrac{(-1)^n}{(2n+1)!} \left(\dfrac{-i\zeta}{2}\right)^{2n} \dfrac{d^{2n}}{d\eta^{2n}} \sinh\left[\frac{\mu\omega}{\hbar}\eta\zeta\right] \nonumber \\
=& \sum_{n=0}^\infty \dfrac{1}{(2n+1)!} \left( \dfrac{\mu\omega}{2\hbar}\zeta^2 \right)^{2n} \tilde{T}^{W}(\eta,\zeta) \nonumber \\
=& \left( \dfrac{\mu\omega}{2\hbar}\zeta^2 \right)^{-1}  \sinh \left[ \dfrac{\mu \omega}{2\hbar}\zeta^2 \right] \tilde{T}^{(W)}(\eta,\zeta),
\end{align}
and is equal to Eq. \eqref{eq:tkfharmonicbj}. Similarly, applying the deformation operator $\mathsf{\hat{\Theta}_{SS}}$ into Eq. \eqref{eq:tkfharmonicweyl}
\begin{align}
\tilde{T}^{SS}(\eta,\zeta) =& \dfrac{1}{2} \left( \dfrac{\mu \omega}{\hbar} \zeta \right)^{-1}  \sum_{n=0}^\infty \dfrac{(-1)^n}{(2n)!} \left(\dfrac{-i\zeta}{2}\right)^{2n} \dfrac{d^{2n}}{d\eta^{2n}}  \sinh\left[\frac{\mu\omega}{\hbar}\eta\zeta\right] \nonumber \\
=& \sum_{n=0}^\infty \dfrac{1}{(2n)!} \left( \dfrac{\mu\omega}{2\hbar}\zeta^2 \right)^{2n} \tilde{T}^{W}(\eta,\zeta) \nonumber \\
=&  \cosh \left[ \dfrac{\mu \omega}{2\hbar}\zeta^2 \right] \tilde{T}^{(W)}(\eta,\zeta),
\end{align}
and again, is equal to Eq. \eqref{eq:tkfharmonicsymmetric}.

\section{The barrier TOA-operator $\mathsf{\hat{T}_B}$}
\label{app:verify}

We now calculate the operator form of the barrier TOA-operator $\mathsf{\hat{T}_B}$ corresponding to the TKF $\tilde{T}_{B,III}(\eta,\zeta)$. Using Eq. \eqref{eq:rhstoa}, we can write the momentum-space representation of $\mathsf{\hat{T}_B}$ as 
\begin{align}
(\mathsf{\hat{T}_B}\phi)(p) = & \int dq\int dq'\int dp'\langle p|q\rangle\langle q|\mathsf{\hat{T}}|q'\rangle\langle q'|p'\rangle\langle p'|\phi\rangle\\
= & \dfrac{\mu}{i\hbar}\int_{-\infty}^{\infty}d\zeta\text{sgn}\zeta\int_{-\infty}^{\infty}dp'\phi(p')\exp\left[-\dfrac{i}{\hbar}(p+p')\frac{\zeta}{2}\right]\int_{-\infty}^{\infty}\dfrac{d\eta}{2\pi\hbar}\exp\left[-\dfrac{i}{\hbar}(p-p')\eta\right]\tilde{T}_{B,III}(\eta,\zeta),
\label{eq:TBopr1}
\end{align}
where, the second line follows from performing a change of variable from the $(q,q')$ to $(\eta,\zeta)$-coordinates, and the plane wave expansion $\braket{q}{p}=e^{iqp/\hbar}/\sqrt{2\pi\hbar}$. Using Eq. \eqref{eq:barrierTKFweyl3}, the integral over $\eta$ in Eq. \eqref{eq:TBopr1} evaluates to 
\begin{align}
\int_{-\infty}^{\infty}\dfrac{d\eta}{2\pi\hbar} \exp\left[-\dfrac{i}{\hbar}(p-p')\eta\right]\tilde{T}_{B,III}(\eta,\zeta) = -\dfrac{i\hbar}{2}\dfrac{d \delta(p-p')}{dp'}+\dfrac{L}{2}\left(1-J_{0}(\kappa_{o}|\zeta|)\right)\delta(p-p').
\label{eq:TBopr2}
\end{align}
Substituting Eq. \eqref{eq:TBopr2} into Eq. \eqref{eq:TBopr1}, we can then evaluate the integral over over $p'$ using integration by parts, which yields
\begin{align}
(\mathsf{\hat{T}_B}\phi)(p) = & \dfrac{\mu}{2i\hbar}\int_{-\infty}^{\infty}d\zeta   \left\{ i\hbar\dfrac{d\phi(p)}{dp}+\frac{\zeta}{2}\phi(p) + L \left(1-J_{0}(\kappa_{o}|\zeta|)\right)\phi(p) \right\} e^{-ip\zeta/\hbar}\text{sgn}\zeta . 
\end{align}
Last, we use the known Fourier transform \cite{gradshteyn2014table}
\begin{align}
\int_{-\infty}^\infty x^{-m} e^{i\sigma x} dx = \dfrac{\pi i^m}{(m-1)!}\sigma^{m-1} \text{sgn}\sigma \qquad m = 1, 2, 3, \dots
\end{align}
and the integral identity Eq. \eqref{eq:integral} to obtain
\begin{align}
(\mathsf{\hat{T}_B}\phi)(p) = -\dfrac{\mu}{2}\left[\dfrac{1}{p}\left(i\hbar\dfrac{d}{dp}+L\right)+\left(i\hbar\dfrac{d}{dp}+L\right)\dfrac{1}{p}\right]\phi(p) +\dfrac{\mu L}{p}\dfrac{H\left(|p|-\sqrt{2\mu V_{o}}\right)}{\sqrt{1-\dfrac{2\mu V_{o}}{p^{2}}}}\phi(p). 
\end{align}
Thus, the operator form of $\mathsf{\hat{T}_B}$ is
\begin{align}
\mathsf{\hat{T}_B} =& -\dfrac{\mu}{2} \bigg( \mathsf{\hat{p}^{-1}} ( \mathsf{\hat{q}} + L ) + ( \mathsf{\hat{q}} + L ) \mathsf{\hat{p}^{-1}}  \bigg) + \mu L \mathsf{\hat{p}^{-1}} ( 1-2\mu V_o \mathsf{\hat{p}^{-2}}  )^{-1/2} H( |\mathsf{\hat{p}}| - \sqrt{2\mu V_o}). 
\end{align}

\end{widetext}

\bibliography{origin-instant.bib}

\begin{thebibliography}{86}%
\makeatletter
\providecommand \@ifxundefined [1]{%
 \@ifx{#1\undefined}
}%
\providecommand \@ifnum [1]{%
 \ifnum #1\expandafter \@firstoftwo
 \else \expandafter \@secondoftwo
 \fi
}%
\providecommand \@ifx [1]{%
 \ifx #1\expandafter \@firstoftwo
 \else \expandafter \@secondoftwo
 \fi
}%
\providecommand \natexlab [1]{#1}%
\providecommand \enquote  [1]{``#1''}%
\providecommand \bibnamefont  [1]{#1}%
\providecommand \bibfnamefont [1]{#1}%
\providecommand \citenamefont [1]{#1}%
\providecommand \href@noop [0]{\@secondoftwo}%
\providecommand \href [0]{\begingroup \@sanitize@url \@href}%
\providecommand \@href[1]{\@@startlink{#1}\@@href}%
\providecommand \@@href[1]{\endgroup#1\@@endlink}%
\providecommand \@sanitize@url [0]{\catcode `\\12\catcode `\$12\catcode
  `\&12\catcode `\#12\catcode `\^12\catcode `\_12\catcode `\%12\relax}%
\providecommand \@@startlink[1]{}%
\providecommand \@@endlink[0]{}%
\providecommand \url  [0]{\begingroup\@sanitize@url \@url }%
\providecommand \@url [1]{\endgroup\@href {#1}{\urlprefix }}%
\providecommand \urlprefix  [0]{URL }%
\providecommand \Eprint [0]{\href }%
\providecommand \doibase [0]{https://doi.org/}%
\providecommand \selectlanguage [0]{\@gobble}%
\providecommand \bibinfo  [0]{\@secondoftwo}%
\providecommand \bibfield  [0]{\@secondoftwo}%
\providecommand \translation [1]{[#1]}%
\providecommand \BibitemOpen [0]{}%
\providecommand \bibitemStop [0]{}%
\providecommand \bibitemNoStop [0]{.\EOS\space}%
\providecommand \EOS [0]{\spacefactor3000\relax}%
\providecommand \BibitemShut  [1]{\csname bibitem#1\endcsname}%
\let\auto@bib@innerbib\@empty
\bibitem [{\citenamefont {MacColl}(1932)}]{maccoll1932note}%
  \BibitemOpen
  \bibfield  {author} {\bibinfo {author} {\bibfnamefont {L.}~\bibnamefont
  {MacColl}},\ }\href@noop {} {\bibfield  {journal} {\bibinfo  {journal}
  {Physical Review}\ }\textbf {\bibinfo {volume} {40}},\ \bibinfo {pages} {621}
  (\bibinfo {year} {1932})}\BibitemShut {NoStop}%
\bibitem [{\citenamefont {Hartman}(1962)}]{hartman1962tunneling}%
  \BibitemOpen
  \bibfield  {author} {\bibinfo {author} {\bibfnamefont {T.~E.}\ \bibnamefont
  {Hartman}},\ }\href@noop {} {\bibfield  {journal} {\bibinfo  {journal}
  {Journal of Applied Physics}\ }\textbf {\bibinfo {volume} {33}},\ \bibinfo
  {pages} {3427} (\bibinfo {year} {1962})}\BibitemShut {NoStop}%
\bibitem [{\citenamefont {Pauli}\ \emph {et~al.}(1933)\citenamefont {Pauli}
  \emph {et~al.}}]{pauli1933handbuch}%
  \BibitemOpen
  \bibfield  {author} {\bibinfo {author} {\bibfnamefont {W.}~\bibnamefont
  {Pauli}} \emph {et~al.},\ }\href@noop {} {\bibfield  {journal} {\bibinfo
  {journal} {Geiger and scheel}\ }\textbf {\bibinfo {volume} {2}},\ \bibinfo
  {pages} {83} (\bibinfo {year} {1933})}\BibitemShut {NoStop}%
\bibitem [{\citenamefont {Wigner}(1955)}]{wigner1955lower}%
  \BibitemOpen
  \bibfield  {author} {\bibinfo {author} {\bibfnamefont {E.~P.}\ \bibnamefont
  {Wigner}},\ }\href@noop {} {\bibfield  {journal} {\bibinfo  {journal}
  {Physical Review}\ }\textbf {\bibinfo {volume} {98}},\ \bibinfo {pages} {145}
  (\bibinfo {year} {1955})}\BibitemShut {NoStop}%
\bibitem [{\citenamefont {B{\"u}ttiker}\ and\ \citenamefont
  {Landauer}(1982)}]{buttiker1982traversal}%
  \BibitemOpen
  \bibfield  {author} {\bibinfo {author} {\bibfnamefont {M.}~\bibnamefont
  {B{\"u}ttiker}}\ and\ \bibinfo {author} {\bibfnamefont {R.}~\bibnamefont
  {Landauer}},\ }\href@noop {} {\bibfield  {journal} {\bibinfo  {journal}
  {Physical Review Letters}\ }\textbf {\bibinfo {volume} {49}},\ \bibinfo
  {pages} {1739} (\bibinfo {year} {1982})}\BibitemShut {NoStop}%
\bibitem [{\citenamefont {Baz}(1966)}]{baz1966lifetime}%
  \BibitemOpen
  \bibfield  {author} {\bibinfo {author} {\bibfnamefont {A.}~\bibnamefont
  {Baz}},\ }\href@noop {} {\bibfield  {journal} {\bibinfo  {journal} {Yadern.
  Fiz.}\ }\textbf {\bibinfo {volume} {4}} (\bibinfo {year} {1966})}\BibitemShut
  {NoStop}%
\bibitem [{\citenamefont {Rybachenko}(1967)}]{rybachenko1967time}%
  \BibitemOpen
  \bibfield  {author} {\bibinfo {author} {\bibfnamefont {V.}~\bibnamefont
  {Rybachenko}},\ }\href@noop {} {\bibfield  {journal} {\bibinfo  {journal}
  {Sov. J. Nucl. Phys.}\ }\textbf {\bibinfo {volume} {5}},\ \bibinfo {pages}
  {635} (\bibinfo {year} {1967})}\BibitemShut {NoStop}%
\bibitem [{\citenamefont {B{\"u}ttiker}(1983)}]{buttiker1983larmor}%
  \BibitemOpen
  \bibfield  {author} {\bibinfo {author} {\bibfnamefont {M.}~\bibnamefont
  {B{\"u}ttiker}},\ }\href@noop {} {\bibfield  {journal} {\bibinfo  {journal}
  {Physical Review B}\ }\textbf {\bibinfo {volume} {27}},\ \bibinfo {pages}
  {6178} (\bibinfo {year} {1983})}\BibitemShut {NoStop}%
\bibitem [{\citenamefont {Pollak}\ and\ \citenamefont
  {Miller}(1984)}]{pollak1984new}%
  \BibitemOpen
  \bibfield  {author} {\bibinfo {author} {\bibfnamefont {E.}~\bibnamefont
  {Pollak}}\ and\ \bibinfo {author} {\bibfnamefont {W.~H.}\ \bibnamefont
  {Miller}},\ }\href@noop {} {\bibfield  {journal} {\bibinfo  {journal}
  {Physical review letters}\ }\textbf {\bibinfo {volume} {53}},\ \bibinfo
  {pages} {115} (\bibinfo {year} {1984})}\BibitemShut {NoStop}%
\bibitem [{\citenamefont {Smith}(1960)}]{smith1960lifetime}%
  \BibitemOpen
  \bibfield  {author} {\bibinfo {author} {\bibfnamefont {F.~T.}\ \bibnamefont
  {Smith}},\ }\href@noop {} {\bibfield  {journal} {\bibinfo  {journal}
  {Physical Review}\ }\textbf {\bibinfo {volume} {118}},\ \bibinfo {pages}
  {349} (\bibinfo {year} {1960})}\BibitemShut {NoStop}%
\bibitem [{\citenamefont {Sokolovski}\ and\ \citenamefont
  {Baskin}(1987)}]{sokolovski1987traversal}%
  \BibitemOpen
  \bibfield  {author} {\bibinfo {author} {\bibfnamefont {D.}~\bibnamefont
  {Sokolovski}}\ and\ \bibinfo {author} {\bibfnamefont {L.}~\bibnamefont
  {Baskin}},\ }\href@noop {} {\bibfield  {journal} {\bibinfo  {journal}
  {Physical Review A}\ }\textbf {\bibinfo {volume} {36}},\ \bibinfo {pages}
  {4604} (\bibinfo {year} {1987})}\BibitemShut {NoStop}%
\bibitem [{\citenamefont {de~Carvalho}\ and\ \citenamefont
  {Nussenzveig}(2002)}]{de2002time}%
  \BibitemOpen
  \bibfield  {author} {\bibinfo {author} {\bibfnamefont {C.~A.}\ \bibnamefont
  {de~Carvalho}}\ and\ \bibinfo {author} {\bibfnamefont {H.~M.}\ \bibnamefont
  {Nussenzveig}},\ }\href@noop {} {\bibfield  {journal} {\bibinfo  {journal}
  {Physics Reports}\ }\textbf {\bibinfo {volume} {364}},\ \bibinfo {pages} {83}
  (\bibinfo {year} {2002})}\BibitemShut {NoStop}%
\bibitem [{\citenamefont {Winful}(2006)}]{winful2006tunneling}%
  \BibitemOpen
  \bibfield  {author} {\bibinfo {author} {\bibfnamefont {H.~G.}\ \bibnamefont
  {Winful}},\ }\href@noop {} {\bibfield  {journal} {\bibinfo  {journal}
  {Physics Reports}\ }\textbf {\bibinfo {volume} {436}},\ \bibinfo {pages} {1}
  (\bibinfo {year} {2006})}\BibitemShut {NoStop}%
\bibitem [{\citenamefont {Imafuku}\ \emph {et~al.}(1997)\citenamefont
  {Imafuku}, \citenamefont {Ohba},\ and\ \citenamefont
  {Yamanaka}}]{imafuku1997effects}%
  \BibitemOpen
  \bibfield  {author} {\bibinfo {author} {\bibfnamefont {K.}~\bibnamefont
  {Imafuku}}, \bibinfo {author} {\bibfnamefont {I.}~\bibnamefont {Ohba}},\ and\
  \bibinfo {author} {\bibfnamefont {Y.}~\bibnamefont {Yamanaka}},\ }\href@noop
  {} {\bibfield  {journal} {\bibinfo  {journal} {Physical Review A}\ }\textbf
  {\bibinfo {volume} {56}},\ \bibinfo {pages} {1142} (\bibinfo {year}
  {1997})}\BibitemShut {NoStop}%
\bibitem [{\citenamefont {Brouard}\ \emph {et~al.}(1994)\citenamefont
  {Brouard}, \citenamefont {Sala},\ and\ \citenamefont
  {Muga}}]{brouard1994systematic}%
  \BibitemOpen
  \bibfield  {author} {\bibinfo {author} {\bibfnamefont {S.}~\bibnamefont
  {Brouard}}, \bibinfo {author} {\bibfnamefont {R.}~\bibnamefont {Sala}},\ and\
  \bibinfo {author} {\bibfnamefont {J.}~\bibnamefont {Muga}},\ }\href@noop {}
  {\bibfield  {journal} {\bibinfo  {journal} {Physical Review A}\ }\textbf
  {\bibinfo {volume} {49}},\ \bibinfo {pages} {4312} (\bibinfo {year}
  {1994})}\BibitemShut {NoStop}%
\bibitem [{\citenamefont {Jaworski}\ and\ \citenamefont
  {Wardlaw}(1988)}]{jaworski1988time}%
  \BibitemOpen
  \bibfield  {author} {\bibinfo {author} {\bibfnamefont {W.}~\bibnamefont
  {Jaworski}}\ and\ \bibinfo {author} {\bibfnamefont {D.~M.}\ \bibnamefont
  {Wardlaw}},\ }\href@noop {} {\bibfield  {journal} {\bibinfo  {journal}
  {Physical Review A}\ }\textbf {\bibinfo {volume} {38}},\ \bibinfo {pages}
  {5404} (\bibinfo {year} {1988})}\BibitemShut {NoStop}%
\bibitem [{\citenamefont {Leavens}\ and\ \citenamefont
  {Aers}(1989)}]{leavens1989dwell}%
  \BibitemOpen
  \bibfield  {author} {\bibinfo {author} {\bibfnamefont {C.}~\bibnamefont
  {Leavens}}\ and\ \bibinfo {author} {\bibfnamefont {G.}~\bibnamefont {Aers}},\
  }\href@noop {} {\bibfield  {journal} {\bibinfo  {journal} {Physical Review
  B}\ }\textbf {\bibinfo {volume} {39}},\ \bibinfo {pages} {1202} (\bibinfo
  {year} {1989})}\BibitemShut {NoStop}%
\bibitem [{\citenamefont {Hauge}\ \emph {et~al.}(1987)\citenamefont {Hauge},
  \citenamefont {Falck},\ and\ \citenamefont
  {Fjeldly}}]{hauge1987transmission}%
  \BibitemOpen
  \bibfield  {author} {\bibinfo {author} {\bibfnamefont {E.}~\bibnamefont
  {Hauge}}, \bibinfo {author} {\bibfnamefont {J.}~\bibnamefont {Falck}},\ and\
  \bibinfo {author} {\bibfnamefont {T.}~\bibnamefont {Fjeldly}},\ }\href@noop
  {} {\bibfield  {journal} {\bibinfo  {journal} {Physical Review B}\ }\textbf
  {\bibinfo {volume} {36}},\ \bibinfo {pages} {4203} (\bibinfo {year}
  {1987})}\BibitemShut {NoStop}%
\bibitem [{\citenamefont {Yamada}(2004)}]{yamada2004unified}%
  \BibitemOpen
  \bibfield  {author} {\bibinfo {author} {\bibfnamefont {N.}~\bibnamefont
  {Yamada}},\ }\href@noop {} {\bibfield  {journal} {\bibinfo  {journal}
  {Physical review letters}\ }\textbf {\bibinfo {volume} {93}},\ \bibinfo
  {pages} {170401} (\bibinfo {year} {2004})}\BibitemShut {NoStop}%
\bibitem [{\citenamefont {Spierings}\ and\ \citenamefont
  {Steinberg}(2021)}]{PhysRevLett.127.133001}%
  \BibitemOpen
  \bibfield  {author} {\bibinfo {author} {\bibfnamefont {D.~C.}\ \bibnamefont
  {Spierings}}\ and\ \bibinfo {author} {\bibfnamefont {A.~M.}\ \bibnamefont
  {Steinberg}},\ }\href {https://doi.org/10.1103/PhysRevLett.127.133001}
  {\bibfield  {journal} {\bibinfo  {journal} {Phys. Rev. Lett.}\ }\textbf
  {\bibinfo {volume} {127}},\ \bibinfo {pages} {133001} (\bibinfo {year}
  {2021})}\BibitemShut {NoStop}%
\bibitem [{\citenamefont {Ara{\'u}jo}\ \emph {et~al.}(2024)\citenamefont
  {Ara{\'u}jo}, \citenamefont {Ximenes},\ and\ \citenamefont
  {Dias}}]{araujo2024space}%
  \BibitemOpen
  \bibfield  {author} {\bibinfo {author} {\bibfnamefont {R.~E.}\ \bibnamefont
  {Ara{\'u}jo}}, \bibinfo {author} {\bibfnamefont {R.}~\bibnamefont
  {Ximenes}},\ and\ \bibinfo {author} {\bibfnamefont {E.~O.}\ \bibnamefont
  {Dias}},\ }\href@noop {} {\bibfield  {journal} {\bibinfo  {journal} {Physical
  Review A}\ }\textbf {\bibinfo {volume} {109}},\ \bibinfo {pages} {012221}
  (\bibinfo {year} {2024})}\BibitemShut {NoStop}%
\bibitem [{\citenamefont {de~Lara}\ and\ \citenamefont
  {Beims}(2024)}]{de2024traveling}%
  \BibitemOpen
  \bibfield  {author} {\bibinfo {author} {\bibfnamefont {A.~J.}\ \bibnamefont
  {de~Lara}}\ and\ \bibinfo {author} {\bibfnamefont {M.~W.}\ \bibnamefont
  {Beims}},\ }\href@noop {} {\bibfield  {journal} {\bibinfo  {journal}
  {Physical Review A}\ }\textbf {\bibinfo {volume} {110}},\ \bibinfo {pages}
  {012216} (\bibinfo {year} {2024})}\BibitemShut {NoStop}%
\bibitem [{\citenamefont {Winful}(2003)}]{winful2003nature}%
  \BibitemOpen
  \bibfield  {author} {\bibinfo {author} {\bibfnamefont {H.~G.}\ \bibnamefont
  {Winful}},\ }\href@noop {} {\bibfield  {journal} {\bibinfo  {journal}
  {Physical review letters}\ }\textbf {\bibinfo {volume} {90}},\ \bibinfo
  {pages} {023901} (\bibinfo {year} {2003})}\BibitemShut {NoStop}%
\bibitem [{\citenamefont {Steinberg}\ \emph {et~al.}(1993)\citenamefont
  {Steinberg}, \citenamefont {Kwiat},\ and\ \citenamefont
  {Chiao}}]{steinberg1993measurement}%
  \BibitemOpen
  \bibfield  {author} {\bibinfo {author} {\bibfnamefont {A.~M.}\ \bibnamefont
  {Steinberg}}, \bibinfo {author} {\bibfnamefont {P.~G.}\ \bibnamefont
  {Kwiat}},\ and\ \bibinfo {author} {\bibfnamefont {R.~Y.}\ \bibnamefont
  {Chiao}},\ }\href
  {https://journals.aps.org/prl/abstract/10.1103/PhysRevLett.71.708} {\bibfield
   {journal} {\bibinfo  {journal} {Physical Review Letters}\ }\textbf {\bibinfo
  {volume} {71}},\ \bibinfo {pages} {708} (\bibinfo {year} {1993})}\BibitemShut
  {NoStop}%
\bibitem [{\citenamefont {Chiao}\ and\ \citenamefont
  {Steinberg}(1997)}]{chiao1997vi}%
  \BibitemOpen
  \bibfield  {author} {\bibinfo {author} {\bibfnamefont {R.~Y.}\ \bibnamefont
  {Chiao}}\ and\ \bibinfo {author} {\bibfnamefont {A.~M.}\ \bibnamefont
  {Steinberg}},\ }in\ \href@noop {} {\emph {\bibinfo {booktitle} {Progress in
  Optics}}},\ Vol.~\bibinfo {volume} {37}\ (\bibinfo  {publisher} {Elsevier},\
  \bibinfo {year} {1997})\ pp.\ \bibinfo {pages} {345--405}\BibitemShut
  {NoStop}%
\bibitem [{\citenamefont {Eckle}\ \emph
  {et~al.}(2008{\natexlab{a}})\citenamefont {Eckle}, \citenamefont {Smolarski},
  \citenamefont {Schlup}, \citenamefont {Biegert}, \citenamefont {Staudte},
  \citenamefont {Sch{\"o}ffler}, \citenamefont {Muller}, \citenamefont
  {D{\"o}rner},\ and\ \citenamefont {Keller}}]{eckle2008attosecondb}%
  \BibitemOpen
  \bibfield  {author} {\bibinfo {author} {\bibfnamefont {P.}~\bibnamefont
  {Eckle}}, \bibinfo {author} {\bibfnamefont {M.}~\bibnamefont {Smolarski}},
  \bibinfo {author} {\bibfnamefont {P.}~\bibnamefont {Schlup}}, \bibinfo
  {author} {\bibfnamefont {J.}~\bibnamefont {Biegert}}, \bibinfo {author}
  {\bibfnamefont {A.}~\bibnamefont {Staudte}}, \bibinfo {author} {\bibfnamefont
  {M.}~\bibnamefont {Sch{\"o}ffler}}, \bibinfo {author} {\bibfnamefont {H.~G.}\
  \bibnamefont {Muller}}, \bibinfo {author} {\bibfnamefont {R.}~\bibnamefont
  {D{\"o}rner}},\ and\ \bibinfo {author} {\bibfnamefont {U.}~\bibnamefont
  {Keller}},\ }\href@noop {} {\bibfield  {journal} {\bibinfo  {journal} {Nature
  Physics}\ }\textbf {\bibinfo {volume} {4}},\ \bibinfo {pages} {565} (\bibinfo
  {year} {2008}{\natexlab{a}})}\BibitemShut {NoStop}%
\bibitem [{\citenamefont {Eckle}\ \emph
  {et~al.}(2008{\natexlab{b}})\citenamefont {Eckle}, \citenamefont {Smolarski},
  \citenamefont {Schlup}, \citenamefont {Biegert}, \citenamefont {Staudte},
  \citenamefont {Sch{\"o}ffler}, \citenamefont {Muller}, \citenamefont
  {D{\"o}rner},\ and\ \citenamefont {Keller}}]{eckle2008attosecond}%
  \BibitemOpen
  \bibfield  {author} {\bibinfo {author} {\bibfnamefont {P.}~\bibnamefont
  {Eckle}}, \bibinfo {author} {\bibfnamefont {M.}~\bibnamefont {Smolarski}},
  \bibinfo {author} {\bibfnamefont {P.}~\bibnamefont {Schlup}}, \bibinfo
  {author} {\bibfnamefont {J.}~\bibnamefont {Biegert}}, \bibinfo {author}
  {\bibfnamefont {A.}~\bibnamefont {Staudte}}, \bibinfo {author} {\bibfnamefont
  {M.}~\bibnamefont {Sch{\"o}ffler}}, \bibinfo {author} {\bibfnamefont {H.~G.}\
  \bibnamefont {Muller}}, \bibinfo {author} {\bibfnamefont {R.}~\bibnamefont
  {D{\"o}rner}},\ and\ \bibinfo {author} {\bibfnamefont {U.}~\bibnamefont
  {Keller}},\ }\href {https://www.nature.com/articles/nphys982} {\bibfield
  {journal} {\bibinfo  {journal} {Nature Physics}\ }\textbf {\bibinfo {volume}
  {4}},\ \bibinfo {pages} {565} (\bibinfo {year}
  {2008}{\natexlab{b}})}\BibitemShut {NoStop}%
\bibitem [{\citenamefont {Pfeiffer}\ \emph {et~al.}(2012)\citenamefont
  {Pfeiffer}, \citenamefont {Cirelli}, \citenamefont {Smolarski}, \citenamefont
  {Dimitrovski}, \citenamefont {Abu-Samha}, \citenamefont {Madsen},\ and\
  \citenamefont {Keller}}]{pfeiffer2012attoclock}%
  \BibitemOpen
  \bibfield  {author} {\bibinfo {author} {\bibfnamefont {A.~N.}\ \bibnamefont
  {Pfeiffer}}, \bibinfo {author} {\bibfnamefont {C.}~\bibnamefont {Cirelli}},
  \bibinfo {author} {\bibfnamefont {M.}~\bibnamefont {Smolarski}}, \bibinfo
  {author} {\bibfnamefont {D.}~\bibnamefont {Dimitrovski}}, \bibinfo {author}
  {\bibfnamefont {M.}~\bibnamefont {Abu-Samha}}, \bibinfo {author}
  {\bibfnamefont {L.~B.}\ \bibnamefont {Madsen}},\ and\ \bibinfo {author}
  {\bibfnamefont {U.}~\bibnamefont {Keller}},\ }\href
  {https://www.nature.com/articles/nphys2125} {\bibfield  {journal} {\bibinfo
  {journal} {Nature Physics}\ }\textbf {\bibinfo {volume} {8}},\ \bibinfo
  {pages} {76} (\bibinfo {year} {2012})}\BibitemShut {NoStop}%
\bibitem [{\citenamefont {Pfeiffer}\ \emph {et~al.}(2013)\citenamefont
  {Pfeiffer}, \citenamefont {Cirelli}, \citenamefont {Smolarski},\ and\
  \citenamefont {Keller}}]{pfeiffer2013recent}%
  \BibitemOpen
  \bibfield  {author} {\bibinfo {author} {\bibfnamefont {A.~N.}\ \bibnamefont
  {Pfeiffer}}, \bibinfo {author} {\bibfnamefont {C.}~\bibnamefont {Cirelli}},
  \bibinfo {author} {\bibfnamefont {M.}~\bibnamefont {Smolarski}},\ and\
  \bibinfo {author} {\bibfnamefont {U.}~\bibnamefont {Keller}},\ }\href
  {https://www.sciencedirect.com/science/article/pii/S0301010412000626?casa_token=7l0iNa0xO9wAAAAA:T3_ZKDhPSzbTPQAo5xq_kaWdPsI7sjCY7ZCfjgx2jPqI1iF1Ru1YPY0mKC8Hi5JSoiWD9Ud14g}
  {\bibfield  {journal} {\bibinfo  {journal} {Chemical Physics}\ }\textbf
  {\bibinfo {volume} {414}},\ \bibinfo {pages} {84} (\bibinfo {year}
  {2013})}\BibitemShut {NoStop}%
\bibitem [{\citenamefont {Sainadh}\ \emph {et~al.}(2019)\citenamefont
  {Sainadh}, \citenamefont {Xu}, \citenamefont {Wang}, \citenamefont
  {Atia-Tul-Noor}, \citenamefont {Wallace}, \citenamefont {Douguet},
  \citenamefont {Bray}, \citenamefont {Ivanov}, \citenamefont {Bartschat},
  \citenamefont {Kheifets} \emph {et~al.}}]{sainadh2019attosecond}%
  \BibitemOpen
  \bibfield  {author} {\bibinfo {author} {\bibfnamefont {U.~S.}\ \bibnamefont
  {Sainadh}}, \bibinfo {author} {\bibfnamefont {H.}~\bibnamefont {Xu}},
  \bibinfo {author} {\bibfnamefont {X.}~\bibnamefont {Wang}}, \bibinfo {author}
  {\bibfnamefont {A.}~\bibnamefont {Atia-Tul-Noor}}, \bibinfo {author}
  {\bibfnamefont {W.~C.}\ \bibnamefont {Wallace}}, \bibinfo {author}
  {\bibfnamefont {N.}~\bibnamefont {Douguet}}, \bibinfo {author} {\bibfnamefont
  {A.}~\bibnamefont {Bray}}, \bibinfo {author} {\bibfnamefont {I.}~\bibnamefont
  {Ivanov}}, \bibinfo {author} {\bibfnamefont {K.}~\bibnamefont {Bartschat}},
  \bibinfo {author} {\bibfnamefont {A.}~\bibnamefont {Kheifets}}, \emph
  {et~al.},\ }\href {https://www.nature.com/articles/s41586-019-1028-3}
  {\bibfield  {journal} {\bibinfo  {journal} {Nature}\ }\textbf {\bibinfo
  {volume} {568}},\ \bibinfo {pages} {75} (\bibinfo {year} {2019})}\BibitemShut
  {NoStop}%
\bibitem [{\citenamefont {Landsman}\ \emph {et~al.}(2014)\citenamefont
  {Landsman}, \citenamefont {Weger}, \citenamefont {Maurer}, \citenamefont
  {Boge}, \citenamefont {Ludwig}, \citenamefont {Heuser}, \citenamefont
  {Cirelli}, \citenamefont {Gallmann},\ and\ \citenamefont
  {Keller}}]{landsman2014ultrafast}%
  \BibitemOpen
  \bibfield  {author} {\bibinfo {author} {\bibfnamefont {A.~S.}\ \bibnamefont
  {Landsman}}, \bibinfo {author} {\bibfnamefont {M.}~\bibnamefont {Weger}},
  \bibinfo {author} {\bibfnamefont {J.}~\bibnamefont {Maurer}}, \bibinfo
  {author} {\bibfnamefont {R.}~\bibnamefont {Boge}}, \bibinfo {author}
  {\bibfnamefont {A.}~\bibnamefont {Ludwig}}, \bibinfo {author} {\bibfnamefont
  {S.}~\bibnamefont {Heuser}}, \bibinfo {author} {\bibfnamefont
  {C.}~\bibnamefont {Cirelli}}, \bibinfo {author} {\bibfnamefont
  {L.}~\bibnamefont {Gallmann}},\ and\ \bibinfo {author} {\bibfnamefont
  {U.}~\bibnamefont {Keller}},\ }\href
  {https://opg.optica.org/optica/fulltext.cfm?uri=optica-1-5-343&id=304679}
  {\bibfield  {journal} {\bibinfo  {journal} {Optica}\ }\textbf {\bibinfo
  {volume} {1}},\ \bibinfo {pages} {343} (\bibinfo {year} {2014})}\BibitemShut
  {NoStop}%
\bibitem [{\citenamefont {Camus}\ \emph {et~al.}(2017)\citenamefont {Camus},
  \citenamefont {Yakaboylu}, \citenamefont {Fechner}, \citenamefont {Klaiber},
  \citenamefont {Laux}, \citenamefont {Mi}, \citenamefont {Hatsagortsyan},
  \citenamefont {Pfeifer}, \citenamefont {Keitel},\ and\ \citenamefont
  {Moshammer}}]{camus2017experimental}%
  \BibitemOpen
  \bibfield  {author} {\bibinfo {author} {\bibfnamefont {N.}~\bibnamefont
  {Camus}}, \bibinfo {author} {\bibfnamefont {E.}~\bibnamefont {Yakaboylu}},
  \bibinfo {author} {\bibfnamefont {L.}~\bibnamefont {Fechner}}, \bibinfo
  {author} {\bibfnamefont {M.}~\bibnamefont {Klaiber}}, \bibinfo {author}
  {\bibfnamefont {M.}~\bibnamefont {Laux}}, \bibinfo {author} {\bibfnamefont
  {Y.}~\bibnamefont {Mi}}, \bibinfo {author} {\bibfnamefont {K.~Z.}\
  \bibnamefont {Hatsagortsyan}}, \bibinfo {author} {\bibfnamefont
  {T.}~\bibnamefont {Pfeifer}}, \bibinfo {author} {\bibfnamefont {C.~H.}\
  \bibnamefont {Keitel}},\ and\ \bibinfo {author} {\bibfnamefont
  {R.}~\bibnamefont {Moshammer}},\ }\href
  {https://journals.aps.org/prl/abstract/10.1103/PhysRevLett.119.023201}
  {\bibfield  {journal} {\bibinfo  {journal} {Physical review letters}\
  }\textbf {\bibinfo {volume} {119}},\ \bibinfo {pages} {023201} (\bibinfo
  {year} {2017})}\BibitemShut {NoStop}%
\bibitem [{\citenamefont {Ramos}\ \emph {et~al.}(2020)\citenamefont {Ramos},
  \citenamefont {Spierings}, \citenamefont {Racicot},\ and\ \citenamefont
  {Steinberg}}]{ramos2020measurement}%
  \BibitemOpen
  \bibfield  {author} {\bibinfo {author} {\bibfnamefont {R.}~\bibnamefont
  {Ramos}}, \bibinfo {author} {\bibfnamefont {D.}~\bibnamefont {Spierings}},
  \bibinfo {author} {\bibfnamefont {I.}~\bibnamefont {Racicot}},\ and\ \bibinfo
  {author} {\bibfnamefont {A.~M.}\ \bibnamefont {Steinberg}},\ }\href
  {https://www.nature.com/articles/s41586-020-2490-7} {\bibfield  {journal}
  {\bibinfo  {journal} {Nature}\ }\textbf {\bibinfo {volume} {583}},\ \bibinfo
  {pages} {529} (\bibinfo {year} {2020})}\BibitemShut {NoStop}%
\bibitem [{\citenamefont {Muga}\ \emph {et~al.}(2007)\citenamefont {Muga},
  \citenamefont {Mayato},\ and\ \citenamefont {Egusquiza}}]{muga2007time}%
  \BibitemOpen
  \bibfield  {author} {\bibinfo {author} {\bibfnamefont {G.}~\bibnamefont
  {Muga}}, \bibinfo {author} {\bibfnamefont {R.~S.}\ \bibnamefont {Mayato}},\
  and\ \bibinfo {author} {\bibfnamefont {I.}~\bibnamefont {Egusquiza}},\
  }\href@noop {} {\emph {\bibinfo {title} {Time in quantum mechanics}}},\ Vol.\
  \bibinfo {volume} {734}\ (\bibinfo  {publisher} {Springer Science \& Business
  Media},\ \bibinfo {year} {2007})\BibitemShut {NoStop}%
\bibitem [{\citenamefont {Galapon}(2002{\natexlab{a}})}]{galapon2002pauli}%
  \BibitemOpen
  \bibfield  {author} {\bibinfo {author} {\bibfnamefont {E.}~\bibnamefont
  {Galapon}},\ }\href@noop {} {\bibfield  {journal} {\bibinfo  {journal}
  {Proceedings of the Royal Society of London. Series A: Mathematical, Physical
  and Engineering Sciences}\ }\textbf {\bibinfo {volume} {458}},\ \bibinfo
  {pages} {451} (\bibinfo {year} {2002}{\natexlab{a}})}\BibitemShut {NoStop}%
\bibitem [{\citenamefont {Galapon}(2002{\natexlab{b}})}]{galapon2002self}%
  \BibitemOpen
  \bibfield  {author} {\bibinfo {author} {\bibfnamefont {E.~A.}\ \bibnamefont
  {Galapon}},\ }\href@noop {} {\bibfield  {journal} {\bibinfo  {journal}
  {Proceedings of the Royal Society of London. Series A: Mathematical, Physical
  and Engineering Sciences}\ }\textbf {\bibinfo {volume} {458}},\ \bibinfo
  {pages} {2671} (\bibinfo {year} {2002}{\natexlab{b}})}\BibitemShut {NoStop}%
\bibitem [{\citenamefont {Galapon}\ \emph {et~al.}(2004)\citenamefont
  {Galapon}, \citenamefont {Caballar},\ and\ \citenamefont
  {Bahague~Jr}}]{galapon2004confined}%
  \BibitemOpen
  \bibfield  {author} {\bibinfo {author} {\bibfnamefont {E.~A.}\ \bibnamefont
  {Galapon}}, \bibinfo {author} {\bibfnamefont {R.~F.}\ \bibnamefont
  {Caballar}},\ and\ \bibinfo {author} {\bibfnamefont {R.~T.}\ \bibnamefont
  {Bahague~Jr}},\ }\href@noop {} {\bibfield  {journal} {\bibinfo  {journal}
  {Physical review letters}\ }\textbf {\bibinfo {volume} {93}},\ \bibinfo
  {pages} {180406} (\bibinfo {year} {2004})}\BibitemShut {NoStop}%
\bibitem [{\citenamefont {Galapon}\ \emph {et~al.}(2005)\citenamefont
  {Galapon}, \citenamefont {Caballar},\ and\ \citenamefont
  {Bahague}}]{galapon2005confined}%
  \BibitemOpen
  \bibfield  {author} {\bibinfo {author} {\bibfnamefont {E.~A.}\ \bibnamefont
  {Galapon}}, \bibinfo {author} {\bibfnamefont {R.~F.}\ \bibnamefont
  {Caballar}},\ and\ \bibinfo {author} {\bibfnamefont {R.}~\bibnamefont
  {Bahague}},\ }\href@noop {} {\bibfield  {journal} {\bibinfo  {journal}
  {Physical Review A}\ }\textbf {\bibinfo {volume} {72}},\ \bibinfo {pages}
  {062107} (\bibinfo {year} {2005})}\BibitemShut {NoStop}%
\bibitem [{\citenamefont {Galapon}(2009)}]{galapon2009quantum}%
  \BibitemOpen
  \bibfield  {author} {\bibinfo {author} {\bibfnamefont {E.~A.}\ \bibnamefont
  {Galapon}},\ }\href@noop {} {\bibfield  {journal} {\bibinfo  {journal}
  {Physical Review A}\ }\textbf {\bibinfo {volume} {80}},\ \bibinfo {pages}
  {030102} (\bibinfo {year} {2009})}\BibitemShut {NoStop}%
\bibitem [{\citenamefont {Sombillo}\ and\ \citenamefont
  {Galapon}(2014)}]{sombillo2014quantum}%
  \BibitemOpen
  \bibfield  {author} {\bibinfo {author} {\bibfnamefont {D.~L.}\ \bibnamefont
  {Sombillo}}\ and\ \bibinfo {author} {\bibfnamefont {E.~A.}\ \bibnamefont
  {Galapon}},\ }\href@noop {} {\bibfield  {journal} {\bibinfo  {journal}
  {Physical Review A}\ }\textbf {\bibinfo {volume} {90}},\ \bibinfo {pages}
  {032115} (\bibinfo {year} {2014})}\BibitemShut {NoStop}%
\bibitem [{\citenamefont {Sombillo}\ and\ \citenamefont
  {Galapon}(2016)}]{sombillo2016particle}%
  \BibitemOpen
  \bibfield  {author} {\bibinfo {author} {\bibfnamefont {D.~L.~B.}\
  \bibnamefont {Sombillo}}\ and\ \bibinfo {author} {\bibfnamefont {E.~A.}\
  \bibnamefont {Galapon}},\ }\href@noop {} {\bibfield  {journal} {\bibinfo
  {journal} {Annals of Physics}\ }\textbf {\bibinfo {volume} {364}},\ \bibinfo
  {pages} {261} (\bibinfo {year} {2016})}\BibitemShut {NoStop}%
\bibitem [{\citenamefont {Galapon}\ and\ \citenamefont
  {Magadan}(2018)}]{galapon2018quantizations}%
  \BibitemOpen
  \bibfield  {author} {\bibinfo {author} {\bibfnamefont {E.~A.}\ \bibnamefont
  {Galapon}}\ and\ \bibinfo {author} {\bibfnamefont {J.~J.~P.}\ \bibnamefont
  {Magadan}},\ }\href@noop {} {\bibfield  {journal} {\bibinfo  {journal}
  {Annals of Physics}\ }\textbf {\bibinfo {volume} {397}},\ \bibinfo {pages}
  {278} (\bibinfo {year} {2018})}\BibitemShut {NoStop}%
\bibitem [{\citenamefont {Flores}\ and\ \citenamefont
  {Galapon}(2019)}]{flores2019quantum}%
  \BibitemOpen
  \bibfield  {author} {\bibinfo {author} {\bibfnamefont {P.~C.~M.}\
  \bibnamefont {Flores}}\ and\ \bibinfo {author} {\bibfnamefont {E.~A.}\
  \bibnamefont {Galapon}},\ }\href@noop {} {\bibfield  {journal} {\bibinfo
  {journal} {Physical Review A}\ }\textbf {\bibinfo {volume} {99}},\ \bibinfo
  {pages} {042113} (\bibinfo {year} {2019})}\BibitemShut {NoStop}%
\bibitem [{\citenamefont {Pablico}\ and\ \citenamefont
  {Galapon}(2020)}]{pablico2020quantum}%
  \BibitemOpen
  \bibfield  {author} {\bibinfo {author} {\bibfnamefont {D.~A.~L.}\
  \bibnamefont {Pablico}}\ and\ \bibinfo {author} {\bibfnamefont {E.~A.}\
  \bibnamefont {Galapon}},\ }\href@noop {} {\bibfield  {journal} {\bibinfo
  {journal} {Physical Review A}\ }\textbf {\bibinfo {volume} {101}},\ \bibinfo
  {pages} {022103} (\bibinfo {year} {2020})}\BibitemShut {NoStop}%
\bibitem [{\citenamefont {Galapon}(2012)}]{galapon2012only}%
  \BibitemOpen
  \bibfield  {author} {\bibinfo {author} {\bibfnamefont {E.~A.}\ \bibnamefont
  {Galapon}},\ }\href@noop {} {\bibfield  {journal} {\bibinfo  {journal}
  {Physical review letters}\ }\textbf {\bibinfo {volume} {108}},\ \bibinfo
  {pages} {170402} (\bibinfo {year} {2012})}\BibitemShut {NoStop}%
\bibitem [{\citenamefont {Torlina}\ \emph {et~al.}(2015)\citenamefont
  {Torlina}, \citenamefont {Morales}, \citenamefont {Kaushal}, \citenamefont
  {Ivanov}, \citenamefont {Kheifets}, \citenamefont {Zielinski}, \citenamefont
  {Scrinzi}, \citenamefont {Muller}, \citenamefont {Sukiasyan}, \citenamefont
  {Ivanov} \emph {et~al.}}]{torlina2015interpreting}%
  \BibitemOpen
  \bibfield  {author} {\bibinfo {author} {\bibfnamefont {L.}~\bibnamefont
  {Torlina}}, \bibinfo {author} {\bibfnamefont {F.}~\bibnamefont {Morales}},
  \bibinfo {author} {\bibfnamefont {J.}~\bibnamefont {Kaushal}}, \bibinfo
  {author} {\bibfnamefont {I.}~\bibnamefont {Ivanov}}, \bibinfo {author}
  {\bibfnamefont {A.}~\bibnamefont {Kheifets}}, \bibinfo {author}
  {\bibfnamefont {A.}~\bibnamefont {Zielinski}}, \bibinfo {author}
  {\bibfnamefont {A.}~\bibnamefont {Scrinzi}}, \bibinfo {author} {\bibfnamefont
  {H.~G.}\ \bibnamefont {Muller}}, \bibinfo {author} {\bibfnamefont
  {S.}~\bibnamefont {Sukiasyan}}, \bibinfo {author} {\bibfnamefont
  {M.}~\bibnamefont {Ivanov}}, \emph {et~al.},\ }\href@noop {} {\bibfield
  {journal} {\bibinfo  {journal} {Nature Physics}\ }\textbf {\bibinfo {volume}
  {11}},\ \bibinfo {pages} {503} (\bibinfo {year} {2015})}\BibitemShut
  {NoStop}%
\bibitem [{\citenamefont {Petersen}\ and\ \citenamefont
  {Pollak}(2017)}]{petersen2017tunneling}%
  \BibitemOpen
  \bibfield  {author} {\bibinfo {author} {\bibfnamefont {J.}~\bibnamefont
  {Petersen}}\ and\ \bibinfo {author} {\bibfnamefont {E.}~\bibnamefont
  {Pollak}},\ }\href@noop {} {\bibfield  {journal} {\bibinfo  {journal} {The
  Journal of Physical Chemistry Letters}\ }\textbf {\bibinfo {volume} {8}},\
  \bibinfo {pages} {4017} (\bibinfo {year} {2017})}\BibitemShut {NoStop}%
\bibitem [{\citenamefont {Petersen}\ and\ \citenamefont
  {Pollak}(2018)}]{petersen2018instantaneous}%
  \BibitemOpen
  \bibfield  {author} {\bibinfo {author} {\bibfnamefont {J.}~\bibnamefont
  {Petersen}}\ and\ \bibinfo {author} {\bibfnamefont {E.}~\bibnamefont
  {Pollak}},\ }\href@noop {} {\bibfield  {journal} {\bibinfo  {journal} {The
  Journal of Physical Chemistry A}\ }\textbf {\bibinfo {volume} {122}},\
  \bibinfo {pages} {3563} (\bibinfo {year} {2018})}\BibitemShut {NoStop}%
\bibitem [{\citenamefont {Dias}\ and\ \citenamefont
  {Parisio}(2017)}]{dias2017space}%
  \BibitemOpen
  \bibfield  {author} {\bibinfo {author} {\bibfnamefont {E.~O.}\ \bibnamefont
  {Dias}}\ and\ \bibinfo {author} {\bibfnamefont {F.}~\bibnamefont {Parisio}},\
  }\href@noop {} {\bibfield  {journal} {\bibinfo  {journal} {Physical Review
  A}\ }\textbf {\bibinfo {volume} {95}},\ \bibinfo {pages} {032133} (\bibinfo
  {year} {2017})}\BibitemShut {NoStop}%
\bibitem [{\citenamefont {Flores}\ and\ \citenamefont
  {Galapon}(2022)}]{flores2022relativistic}%
  \BibitemOpen
  \bibfield  {author} {\bibinfo {author} {\bibfnamefont {P.~C.}\ \bibnamefont
  {Flores}}\ and\ \bibinfo {author} {\bibfnamefont {E.~A.}\ \bibnamefont
  {Galapon}},\ }\href@noop {} {\bibfield  {journal} {\bibinfo  {journal}
  {Physical Review A}\ }\textbf {\bibinfo {volume} {105}},\ \bibinfo {pages}
  {062208} (\bibinfo {year} {2022})}\BibitemShut {NoStop}%
\bibitem [{\citenamefont {Flores}\ and\ \citenamefont
  {Galapon}(2023{\natexlab{a}})}]{flores2023instantaneous}%
  \BibitemOpen
  \bibfield  {author} {\bibinfo {author} {\bibfnamefont {P.~C.}\ \bibnamefont
  {Flores}}\ and\ \bibinfo {author} {\bibfnamefont {E.~A.}\ \bibnamefont
  {Galapon}},\ }\href@noop {} {\bibfield  {journal} {\bibinfo  {journal}
  {Europhysics Letters}\ }\textbf {\bibinfo {volume} {141}},\ \bibinfo {pages}
  {10001} (\bibinfo {year} {2023}{\natexlab{a}})}\BibitemShut {NoStop}%
\bibitem [{\citenamefont {Flores}\ and\ \citenamefont
  {Galapon}(2023{\natexlab{b}})}]{flores2023quantized}%
  \BibitemOpen
  \bibfield  {author} {\bibinfo {author} {\bibfnamefont {P.}~\bibnamefont
  {Flores}}\ and\ \bibinfo {author} {\bibfnamefont {E.~A.}\ \bibnamefont
  {Galapon}},\ }\href@noop {} {\bibfield  {journal} {\bibinfo  {journal} {The
  European Physical Journal Plus}\ }\textbf {\bibinfo {volume} {138}},\
  \bibinfo {pages} {1} (\bibinfo {year} {2023}{\natexlab{b}})}\BibitemShut
  {NoStop}%
\bibitem [{\citenamefont {Jacak}(2023)}]{jacak2023forbidden}%
  \BibitemOpen
  \bibfield  {author} {\bibinfo {author} {\bibfnamefont {J.~E.}\ \bibnamefont
  {Jacak}},\ }\href@noop {} {\bibfield  {journal} {\bibinfo  {journal}
  {Physical Review A}\ }\textbf {\bibinfo {volume} {107}},\ \bibinfo {pages}
  {032207} (\bibinfo {year} {2023})}\BibitemShut {NoStop}%
\bibitem [{\citenamefont {Carosso}(2022)}]{carosso2022quantization}%
  \BibitemOpen
  \bibfield  {author} {\bibinfo {author} {\bibfnamefont {A.}~\bibnamefont
  {Carosso}},\ }\href@noop {} {\bibfield  {journal} {\bibinfo  {journal}
  {Studies in History and Philosophy of Science}\ }\textbf {\bibinfo {volume}
  {96}},\ \bibinfo {pages} {35} (\bibinfo {year} {2022})}\BibitemShut {NoStop}%
\bibitem [{\citenamefont {Ali}\ and\ \citenamefont
  {Doebner}(1990)}]{ali1990ordering}%
  \BibitemOpen
  \bibfield  {author} {\bibinfo {author} {\bibfnamefont {S.~T.}\ \bibnamefont
  {Ali}}\ and\ \bibinfo {author} {\bibfnamefont {H.-D.}\ \bibnamefont
  {Doebner}},\ }\href@noop {} {\bibfield  {journal} {\bibinfo  {journal}
  {Physical Review A}\ }\textbf {\bibinfo {volume} {41}},\ \bibinfo {pages}
  {1199} (\bibinfo {year} {1990})}\BibitemShut {NoStop}%
\bibitem [{\citenamefont {Agarwal}\ and\ \citenamefont
  {Wolf}(1970)}]{agarwal1970calculus}%
  \BibitemOpen
  \bibfield  {author} {\bibinfo {author} {\bibfnamefont {G.~S.}\ \bibnamefont
  {Agarwal}}\ and\ \bibinfo {author} {\bibfnamefont {E.}~\bibnamefont {Wolf}},\
  }\href@noop {} {\bibfield  {journal} {\bibinfo  {journal} {Physical Review
  D}\ }\textbf {\bibinfo {volume} {2}},\ \bibinfo {pages} {2161} (\bibinfo
  {year} {1970})}\BibitemShut {NoStop}%
\bibitem [{\citenamefont {Mackey}(1968)}]{mackey1968induced}%
  \BibitemOpen
  \bibfield  {author} {\bibinfo {author} {\bibfnamefont {G.~W.}\ \bibnamefont
  {Mackey}},\ }\href {https://cir.nii.ac.jp/crid/1130282271848740608} {\emph
  {\bibinfo {title} {Induced representations of groups and quantum
  mechanics}}},\ Pubblicazioni della Classe di scienze\ (\bibinfo  {publisher}
  {W.A. Benjamin and Editore Boringhieri},\ \bibinfo {year} {1968})\BibitemShut
  {NoStop}%
\bibitem [{\citenamefont {Mackey}(1989)}]{mackey1978unitary}%
  \BibitemOpen
  \bibfield  {author} {\bibinfo {author} {\bibfnamefont {G.~W.}\ \bibnamefont
  {Mackey}},\ }\href {https://cir.nii.ac.jp/crid/1130282269832702976} {\emph
  {\bibinfo {title} {Unitary group representations in physics, probability, and
  number theory}}},\ Advanced book classics\ (\bibinfo  {publisher}
  {Addison-Wesley Pub. Co.},\ \bibinfo {year} {1989})\BibitemShut {NoStop}%
\bibitem [{\citenamefont {Birkhoff}\ and\ \citenamefont
  {Von~Neumann}(1975)}]{birkhoff1975logic}%
  \BibitemOpen
  \bibfield  {author} {\bibinfo {author} {\bibfnamefont {G.}~\bibnamefont
  {Birkhoff}}\ and\ \bibinfo {author} {\bibfnamefont {J.}~\bibnamefont
  {Von~Neumann}},\ }in\ \href@noop {} {\emph {\bibinfo {booktitle} {The
  Logico-Algebraic Approach to Quantum Mechanics: Volume I: Historical
  Evolution}}}\ (\bibinfo  {publisher} {Springer},\ \bibinfo {year} {1975})\
  pp.\ \bibinfo {pages} {1--26}\BibitemShut {NoStop}%
\bibitem [{\citenamefont {Galapon}(2004)}]{galapon2004shouldn}%
  \BibitemOpen
  \bibfield  {author} {\bibinfo {author} {\bibfnamefont {E.~A.}\ \bibnamefont
  {Galapon}},\ }\href@noop {} {\bibfield  {journal} {\bibinfo  {journal}
  {Journal of mathematical physics}\ }\textbf {\bibinfo {volume} {45}},\
  \bibinfo {pages} {3180} (\bibinfo {year} {2004})}\BibitemShut {NoStop}%
\bibitem [{\citenamefont {De~la Madrid}(2005)}]{de2005role}%
  \BibitemOpen
  \bibfield  {author} {\bibinfo {author} {\bibfnamefont {R.}~\bibnamefont
  {De~la Madrid}},\ }\href@noop {} {\bibfield  {journal} {\bibinfo  {journal}
  {European journal of physics}\ }\textbf {\bibinfo {volume} {26}},\ \bibinfo
  {pages} {287} (\bibinfo {year} {2005})}\BibitemShut {NoStop}%
\bibitem [{\citenamefont {Pablico}\ and\ \citenamefont
  {Galapon}(2023)}]{pablico2022quantum}%
  \BibitemOpen
  \bibfield  {author} {\bibinfo {author} {\bibfnamefont {D.~A.~L.}\
  \bibnamefont {Pablico}}\ and\ \bibinfo {author} {\bibfnamefont {E.~A.}\
  \bibnamefont {Galapon}},\ }\href@noop {} {\bibfield  {journal} {\bibinfo
  {journal} {The European Physical Journal Plus}\ }\textbf {\bibinfo {volume}
  {138}},\ \bibinfo {pages} {1} (\bibinfo {year} {2023})}\BibitemShut {NoStop}%
\bibitem [{\citenamefont {Le{\'o}n}\ \emph {et~al.}(2000)\citenamefont
  {Le{\'o}n}, \citenamefont {Julve}, \citenamefont {Pitanga},\ and\
  \citenamefont {De~Urr{\'\i}es}}]{leon2000time}%
  \BibitemOpen
  \bibfield  {author} {\bibinfo {author} {\bibfnamefont {J.}~\bibnamefont
  {Le{\'o}n}}, \bibinfo {author} {\bibfnamefont {J.}~\bibnamefont {Julve}},
  \bibinfo {author} {\bibfnamefont {P.}~\bibnamefont {Pitanga}},\ and\ \bibinfo
  {author} {\bibfnamefont {F.}~\bibnamefont {De~Urr{\'\i}es}},\ }\href@noop {}
  {\bibfield  {journal} {\bibinfo  {journal} {Physical Review A}\ }\textbf
  {\bibinfo {volume} {61}},\ \bibinfo {pages} {062101} (\bibinfo {year}
  {2000})}\BibitemShut {NoStop}%
\bibitem [{\citenamefont {Peres}(1997)}]{peres1997quantum}%
  \BibitemOpen
  \bibfield  {author} {\bibinfo {author} {\bibfnamefont {A.}~\bibnamefont
  {Peres}},\ }\href@noop {} {\emph {\bibinfo {title} {Quantum theory: concepts
  and methods}}},\ Vol.~\bibinfo {volume} {72}\ (\bibinfo  {publisher}
  {Springer},\ \bibinfo {year} {1997})\BibitemShut {NoStop}%
\bibitem [{\citenamefont {de~Gosson}(2016)}]{Gosson2016}%
  \BibitemOpen
  \bibfield  {author} {\bibinfo {author} {\bibfnamefont {M.~A.}\ \bibnamefont
  {de~Gosson}},\ }\href@noop {} {\bibfield  {journal} {\bibinfo  {journal}
  {Physics Reports}\ }\textbf {\bibinfo {volume} {623}},\ \bibinfo {pages} {1}
  (\bibinfo {year} {2016})}\BibitemShut {NoStop}%
\bibitem [{\citenamefont {De~Gosson}(2006)}]{de2006symplectic}%
  \BibitemOpen
  \bibfield  {author} {\bibinfo {author} {\bibfnamefont {M.~A.}\ \bibnamefont
  {De~Gosson}},\ }\href@noop {} {\emph {\bibinfo {title} {Symplectic geometry
  and quantum mechanics}}},\ Vol.\ \bibinfo {volume} {166}\ (\bibinfo
  {publisher} {Springer Science \& Business Media},\ \bibinfo {year}
  {2006})\BibitemShut {NoStop}%
\bibitem [{\citenamefont {De~Gosson}(2013)}]{de2013born}%
  \BibitemOpen
  \bibfield  {author} {\bibinfo {author} {\bibfnamefont {M.~A.}\ \bibnamefont
  {De~Gosson}},\ }\href@noop {} {\bibfield  {journal} {\bibinfo  {journal}
  {Journal of Physics A: Mathematical and Theoretical}\ }\textbf {\bibinfo
  {volume} {46}},\ \bibinfo {pages} {445301} (\bibinfo {year}
  {2013})}\BibitemShut {NoStop}%
\bibitem [{\citenamefont {Cohen}(1966)}]{cohen1966generalized}%
  \BibitemOpen
  \bibfield  {author} {\bibinfo {author} {\bibfnamefont {L.}~\bibnamefont
  {Cohen}},\ }\href@noop {} {\bibfield  {journal} {\bibinfo  {journal} {Journal
  of Mathematical Physics}\ }\textbf {\bibinfo {volume} {7}},\ \bibinfo {pages}
  {781} (\bibinfo {year} {1966})}\BibitemShut {NoStop}%
\bibitem [{\citenamefont {Domingo}\ and\ \citenamefont
  {Galapon}(2015)}]{domingo2015generalized}%
  \BibitemOpen
  \bibfield  {author} {\bibinfo {author} {\bibfnamefont {H.~B.}\ \bibnamefont
  {Domingo}}\ and\ \bibinfo {author} {\bibfnamefont {E.~A.}\ \bibnamefont
  {Galapon}},\ }\href@noop {} {\bibfield  {journal} {\bibinfo  {journal}
  {Journal of Mathematical Physics}\ }\textbf {\bibinfo {volume} {56}}
  (\bibinfo {year} {2015})}\BibitemShut {NoStop}%
\bibitem [{\citenamefont {Shewell}(1959)}]{shewell1959formation}%
  \BibitemOpen
  \bibfield  {author} {\bibinfo {author} {\bibfnamefont {J.~R.}\ \bibnamefont
  {Shewell}},\ }\href@noop {} {\bibfield  {journal} {\bibinfo  {journal}
  {American Journal of Physics}\ }\textbf {\bibinfo {volume} {27}},\ \bibinfo
  {pages} {16} (\bibinfo {year} {1959})}\BibitemShut {NoStop}%
\bibitem [{\citenamefont {De~Gosson}(2016)}]{de2016born}%
  \BibitemOpen
  \bibfield  {author} {\bibinfo {author} {\bibfnamefont {M.~A.}\ \bibnamefont
  {De~Gosson}},\ }\href@noop {} {\emph {\bibinfo {title} {Born-Jordan
  quantization: theory and applications}}},\ Vol.\ \bibinfo {volume} {182}\
  (\bibinfo  {publisher} {Springer},\ \bibinfo {year} {2016})\BibitemShut
  {NoStop}%
\bibitem [{\citenamefont {Cohen}(2012)}]{cohen2012weyl}%
  \BibitemOpen
  \bibfield  {author} {\bibinfo {author} {\bibfnamefont {L.}~\bibnamefont
  {Cohen}},\ }\href@noop {} {\emph {\bibinfo {title} {The Weyl operator and its
  generalization}}}\ (\bibinfo  {publisher} {Springer Science \& Business
  Media},\ \bibinfo {year} {2012})\BibitemShut {NoStop}%
\bibitem [{\citenamefont {de~Gosson}(2016)}]{Gosson2016a}%
  \BibitemOpen
  \bibfield  {author} {\bibinfo {author} {\bibfnamefont {M.~A.}\ \bibnamefont
  {de~Gosson}},\ }\bibinfo {title} {Born–{{Jordan Quantization}}},\ in\
  \href@noop {} {\emph {\bibinfo {booktitle} {Born-{{Jordan Quantization}}}}},\
  \bibinfo {series} {Fundamental {{Theories}} of {{Physics}}}, Vol.\ \bibinfo
  {volume} {182}\ (\bibinfo  {publisher} {{Springer International
  Publishing}},\ \bibinfo {address} {{Cham}},\ \bibinfo {year} {2016})\ pp.\
  \bibinfo {pages} {113--127}\BibitemShut {NoStop}%
\bibitem [{\citenamefont {de~Gosson}\ and\ \citenamefont
  {Luef}(2011)}]{Gosson2011}%
  \BibitemOpen
  \bibfield  {author} {\bibinfo {author} {\bibfnamefont {M.}~\bibnamefont
  {de~Gosson}}\ and\ \bibinfo {author} {\bibfnamefont {F.}~\bibnamefont
  {Luef}},\ }\href@noop {} {\bibfield  {journal} {\bibinfo  {journal} {Journal
  of Pseudo-Differential Operators and Applications}\ }\textbf {\bibinfo
  {volume} {2}},\ \bibinfo {pages} {115} (\bibinfo {year} {2011})}\BibitemShut
  {NoStop}%
\bibitem [{\citenamefont {Bender}\ and\ \citenamefont
  {Dunne}(1989{\natexlab{a}})}]{bender1989exact}%
  \BibitemOpen
  \bibfield  {author} {\bibinfo {author} {\bibfnamefont {C.~M.}\ \bibnamefont
  {Bender}}\ and\ \bibinfo {author} {\bibfnamefont {G.~V.}\ \bibnamefont
  {Dunne}},\ }\href@noop {} {\bibfield  {journal} {\bibinfo  {journal}
  {Physical Review D}\ }\textbf {\bibinfo {volume} {40}},\ \bibinfo {pages}
  {2739} (\bibinfo {year} {1989}{\natexlab{a}})}\BibitemShut {NoStop}%
\bibitem [{\citenamefont {Bender}\ and\ \citenamefont
  {Dunne}(1989{\natexlab{b}})}]{bender1989integration}%
  \BibitemOpen
  \bibfield  {author} {\bibinfo {author} {\bibfnamefont {C.~M.}\ \bibnamefont
  {Bender}}\ and\ \bibinfo {author} {\bibfnamefont {G.~V.}\ \bibnamefont
  {Dunne}},\ }\href@noop {} {\bibfield  {journal} {\bibinfo  {journal}
  {Physical Review D}\ }\textbf {\bibinfo {volume} {40}},\ \bibinfo {pages}
  {3504} (\bibinfo {year} {1989}{\natexlab{b}})}\BibitemShut {NoStop}%
\bibitem [{\citenamefont {Bender}\ and\ \citenamefont
  {Gianfreda}(2012)}]{bender2012matrix}%
  \BibitemOpen
  \bibfield  {author} {\bibinfo {author} {\bibfnamefont {C.~M.}\ \bibnamefont
  {Bender}}\ and\ \bibinfo {author} {\bibfnamefont {M.}~\bibnamefont
  {Gianfreda}},\ }\href@noop {} {\bibfield  {journal} {\bibinfo  {journal}
  {Journal of mathematical physics}\ }\textbf {\bibinfo {volume} {53}}
  (\bibinfo {year} {2012})}\BibitemShut {NoStop}%
\bibitem [{\citenamefont {Domingo}(2016)}]{domingothesis}%
  \BibitemOpen
  \bibfield  {author} {\bibinfo {author} {\bibfnamefont {H.~B.}\ \bibnamefont
  {Domingo}},\ }\emph {\bibinfo {title} {Algebra of non-Weyl operators and
  quantization of analytic and non-analytic classical functions}},\ \href@noop
  {} {\bibinfo {type} {Phd thesis}},\ \bibinfo  {school} {University of the
  Philippines Diliman} (\bibinfo {year} {2016}),\ \bibinfo {note} {dissertation
  Adviser: Galapon, E.A.}\BibitemShut {Stop}%
\bibitem [{\citenamefont {Gel'fand}\ and\ \citenamefont
  {Shilov}(1969)}]{gelgeneralized}%
  \BibitemOpen
  \bibfield  {author} {\bibinfo {author} {\bibfnamefont {I.}~\bibnamefont
  {Gel'fand}}\ and\ \bibinfo {author} {\bibfnamefont {G.}~\bibnamefont
  {Shilov}},\ }\href@noop {} {\bibfield  {journal} {\bibinfo  {journal}
  {GelfandIGeneralized Functions1964}\ } (\bibinfo {year} {1969})},\ \bibinfo
  {note} {p.360}\BibitemShut {NoStop}%
\bibitem [{\citenamefont {Groenewold}\ and\ \citenamefont
  {Groenewold}(1946)}]{groenewold1946principles}%
  \BibitemOpen
  \bibfield  {author} {\bibinfo {author} {\bibfnamefont {H.~J.}\ \bibnamefont
  {Groenewold}}\ and\ \bibinfo {author} {\bibfnamefont {H.~J.}\ \bibnamefont
  {Groenewold}},\ }\href@noop {} {\emph {\bibinfo {title} {On the principles of
  elementary quantum mechanics}}}\ (\bibinfo  {publisher} {Springer},\ \bibinfo
  {year} {1946})\BibitemShut {NoStop}%
\bibitem [{\citenamefont {Van~Hove}(1951)}]{van1951certaines}%
  \BibitemOpen
  \bibfield  {author} {\bibinfo {author} {\bibfnamefont {L.~C.~P.}\
  \bibnamefont {Van~Hove}},\ }\emph {\bibinfo {title} {Sur certaines
  repr{\'e}sentations unitaires d'un groupe infini de transformations}},\
  \href@noop {} {Ph.D. thesis},\ \bibinfo  {school} {Bruxelles U.} (\bibinfo
  {year} {1951})\BibitemShut {NoStop}%
\bibitem [{\citenamefont {Gotay}(1999)}]{gotay1999groenewold}%
  \BibitemOpen
  \bibfield  {author} {\bibinfo {author} {\bibfnamefont {M.~J.}\ \bibnamefont
  {Gotay}},\ }\href@noop {} {\bibfield  {journal} {\bibinfo  {journal} {Journal
  of Mathematical Physics}\ }\textbf {\bibinfo {volume} {40}},\ \bibinfo
  {pages} {2107} (\bibinfo {year} {1999})}\BibitemShut {NoStop}%
\bibitem [{\citenamefont {Aharonov}\ and\ \citenamefont
  {Bohm}(1961)}]{aharonov1961time}%
  \BibitemOpen
  \bibfield  {author} {\bibinfo {author} {\bibfnamefont {Y.}~\bibnamefont
  {Aharonov}}\ and\ \bibinfo {author} {\bibfnamefont {D.}~\bibnamefont
  {Bohm}},\ }\href@noop {} {\bibfield  {journal} {\bibinfo  {journal} {Physical
  Review}\ }\textbf {\bibinfo {volume} {122}},\ \bibinfo {pages} {1649}
  (\bibinfo {year} {1961})}\BibitemShut {NoStop}%
\bibitem [{\citenamefont {Gradshteyn}\ and\ \citenamefont
  {Ryzhik}(2014)}]{gradshteyn2014table}%
  \BibitemOpen
  \bibfield  {author} {\bibinfo {author} {\bibfnamefont {I.~S.}\ \bibnamefont
  {Gradshteyn}}\ and\ \bibinfo {author} {\bibfnamefont {I.~M.}\ \bibnamefont
  {Ryzhik}},\ }\href@noop {} {\emph {\bibinfo {title} {Table of integrals,
  series, and products}}}\ (\bibinfo  {publisher} {Academic press},\ \bibinfo
  {year} {2014})\BibitemShut {NoStop}%
\bibitem [{\citenamefont {Giannitrapani}(1997)}]{giannitrapani1997positive}%
  \BibitemOpen
  \bibfield  {author} {\bibinfo {author} {\bibfnamefont {R.}~\bibnamefont
  {Giannitrapani}},\ }\href@noop {} {\bibfield  {journal} {\bibinfo  {journal}
  {International Journal of Theoretical Physics}\ }\textbf {\bibinfo {volume}
  {36}},\ \bibinfo {pages} {1575} (\bibinfo {year} {1997})}\BibitemShut
  {NoStop}%
\bibitem [{\citenamefont {Flores}\ \emph {et~al.}(2024)\citenamefont {Flores},
  \citenamefont {Pablico},\ and\ \citenamefont {Galapon}}]{flores2024partial}%
  \BibitemOpen
  \bibfield  {author} {\bibinfo {author} {\bibfnamefont {P.~C.}\ \bibnamefont
  {Flores}}, \bibinfo {author} {\bibfnamefont {D.~A.~L.}\ \bibnamefont
  {Pablico}},\ and\ \bibinfo {author} {\bibfnamefont {E.~A.}\ \bibnamefont
  {Galapon}},\ }\href@noop {} {\bibfield  {journal} {\bibinfo  {journal}
  {Europhysics Letters}\ }\textbf {\bibinfo {volume} {145}},\ \bibinfo {pages}
  {65002} (\bibinfo {year} {2024})}\BibitemShut {NoStop}%
\end{thebibliography}%

\end{document}